\documentclass[prx,reprint,superscriptaddress,aps,longbibliography,twocolumn]{revtex4-2} 

\usepackage{graphicx}
\usepackage{float}
\usepackage{dcolumn}
\usepackage{amsmath}					
\usepackage{amsfonts}	 
\usepackage{amsthm}					
\usepackage{mathrsfs} 					
\usepackage{bm}				
\usepackage{bbm}
\usepackage{enumerate} 
\usepackage{physics}
\usepackage{appendix}
\usepackage{graphicx}
\usepackage{color}
\usepackage[dvipsnames]{xcolor}
\usepackage{orcidlink}

\usepackage{array}
\usepackage{tabularx}
\usepackage{booktabs}

\usepackage{siunitx}    
\usepackage{multirow}

\usepackage{hyperref}
\hypersetup{
	colorlinks=true,
	linkcolor=blue,
	citecolor=blue,
	filecolor=magenta,
	urlcolor=blue
}

\makeatletter
\providecommand{\DIFadd}[1]{}
\renewcommand{\DIFadd}[1]{\textcolor{blue}{#1}}
\makeatother

\usepackage{lineno}
\newcommand{\vc}[1]{\bm{\mathrm{#1}}}

\newcommand{\jak}{$\pi\!-\!2\pi\!-\!\pi$ }

\begin{document}


\preprint{APS/123-QED}

\title{High-fidelity non-adiabatic dark state gates for neutral atoms}


\author{Nader Mostaan\,\orcidlink{0000-0002-9573-7608}}
\email{nader.mostaan@uni-hamburg.de}
\affiliation{Zentrum für Optische Quantentechnologien, Universität Hamburg, Luruper Chaussee 149, 22761 Hamburg, Germany}

\author{Kapil Goswami\,\orcidlink{0009-0007-7282-7970}}
\affiliation{Zentrum für Optische Quantentechnologien, Universität Hamburg, Luruper Chaussee 149, 22761 Hamburg, Germany}

\author{Peter Schmelcher\,\orcidlink{0000-0002-2637-0937}}
\affiliation{Zentrum für Optische Quantentechnologien, Universität Hamburg, Luruper Chaussee 149, 22761 Hamburg, Germany}
\affiliation{The Hamburg Centre for Ultrafast Imaging, Universität Hamburg, Luruper Chaussee 149, 22761 Hamburg, Germany}

\author{Rick Mukherjee\,\orcidlink{0000-0001-9267-4421}}
\affiliation{Zentrum für Optische Quantentechnologien, Universität Hamburg, Luruper Chaussee 149, 22761 Hamburg, Germany}
\affiliation{Department of Physics $\&$ Astronomy, University of Tennessee, Chattanooga, TN 37403, USA}
\affiliation{UTC Quantum Center, University of Tennessee, Chattanooga, TN 37403, USA}

\date{\today}



\begin{abstract}

Rydberg blockade gates are the most experimentally mature entangling operations in neutral-atom quantum processors, combining fast gate times with simple control, but their performance degrades at larger interatomic separations and remains sensitive to motional and technical noise. Non-blockade gate schemes, such as dark state and geometric protocols, offer complementary robustness but typically rely on complex and experimentally demanding control. Here we show that quantum optimal control enables non-blockade gate schemes to be implemented using the experimentally established pulse-shaping techniques developed for blockade-based gates. Focusing on the dark state gate, we construct non-adiabatic implementations that preserve the intrinsic robustness of adiabatic dark state protocols while achieving gate times comparable to time-optimal blockade gates using only smooth, experimentally feasible pulses. The resulting gates exhibit enhanced resilience to motional coupling, laser noise, and interaction inhomogeneity, particularly near and beyond the blockade radius. This work establishes a practical route to fast, robust non-blockade-based two-qubit gates without increased experimental complexity.

\end{abstract}


\maketitle

\section{Introduction}

Gate-based quantum computation has emerged as a leading paradigm in quantum science and technology, offering significant potential for applications such as the simulation of complex quantum materials~\cite{tazhigulov2022simulating,cao2023ab,clinton2024towards,ma2020quantum}, solving classically intractable optimization problems~\cite{farhi2014quantum,zhou2020quantum,wang2018quantum,goswami2025qudit,koch2025resource}, and design of novel molecules and pharmaceuticals~\cite{li2024hybrid,blunt2022perspective,smaldone2025quantum}.

The ability to perform quantum information processing tasks in gate-based quantum devices requires the precise control of individual qubits and implementation of high-fidelity entangling operations~\cite{barenco1995elementary,divincenzo2000physical,ladd2010quantum}. Such control enables quantum algorithms~\cite{Montanaro2016}, error correction~\cite{terhal2015quantum,katabarwa2024early}, and the generation of complex entangled states~\cite{neeley2010generation}, which are essential for realizing practical quantum technologies. In recent years, significant progress has been made across several physical platforms, including superconducting circuits \cite{arute2019quantum}, trapped ions~\cite{nunnerich2025fast}, photonic systems~\cite{lita2021quantum}, and neutral atoms~\cite{henriet2020quantum,zhou2025low}, each offering complementary advantages for scalability and coherence~\cite{brown2016co}. In all these platforms, the fidelity of entangling gates directly impacts the performance of algorithms and the feasibility of fault-tolerant architectures, motivating extensive research into optimizing gate protocols across different qubit platforms~\cite{nunnerich2025fast,jandura2023optimizing,glaser2025closed,egorova2025three}.

Significant efforts have been devoted to designing high-fidelity multi-qubit and two-qubit gates in a variety of physical systems. In superconducting circuits, controlled-NOT (CNOT) and CZ gates have been realized with fidelities exceeding 0.99 using microwave-driven interactions~\cite{lin202524,ye2021realization,barends2014superconducting,kandala2019error}. Photonic systems have demonstrated probabilistic entangling gates through linear optics and measurement-induced interactions~\cite{KLM2001,kok2007linear}. 

In neutral atom platforms, two-qubit gates have been implemented using strong interactions between highly excited Rydberg states~\cite{saffman2010quantum,Levine2019,Graham2019}. Neutral atom systems offer unique advantages for scalable quantum computation, due to various factors, among them long coherence times \cite{saffman2010quantum}, flexible and reconfigurable arrays in different dimensions~\cite{Bluvstein2022,Bluvstein2024,Evered2023} and long-range, controllable interactions via Rydberg states, to name a few. Individual atoms can be trapped in optical tweezers with high spatial resolution and rearranged into arbitrary geometries~\cite{endres2016atom}, while Rydberg excitations provide strong, controllable interactions that enable fast entangling gates while maintaining long coherence times~\cite{saffman2010quantum,Levine2019}.

Implementing high-fidelity two-qubit gates in neutral atom systems is particularly significant, as they are the fundamental building blocks for multi-qubit operations and scalable quantum algorithms. The main goals are to achieve fast gate operations, minimize residual entanglement with motional degrees of freedom, and maintain robustness against technical noise. Controlled-$Z$ (CZ) gates based on Rydberg interactions have been demonstrated experimentally with fidelities approaching the $0.995$–$0.996$ range~\cite{evered2023high,lin2026sustaining}, and reaching up to $0.998$ when combined with non-destructive erasure detection~\cite{lin2026sustaining}. These error rates already achieve the thresholds of certain quantum error correcting codes, such as the $XZZX$ surface code, which exhibits relatively high tolerance to noise~\cite{sahay2023high}. The precise threshold depends on the underlying error model and architectural assumptions. For instance, values of $8.2\%$ (for biased erasure error) and $4.3\%$ (for unbiased erasure error) have been reported when $98\%$ of two-qubit gate errors correspond to biased erasures (see Table I of Ref.~\cite{sahay2023high}). Nevertheless, further improvements in two-qubit gate fidelity and robustness remain highly desirable, as they directly enhance code performance and reduce resource overheads.

\begin{figure*}
    \centering
    \includegraphics[width=\textwidth]{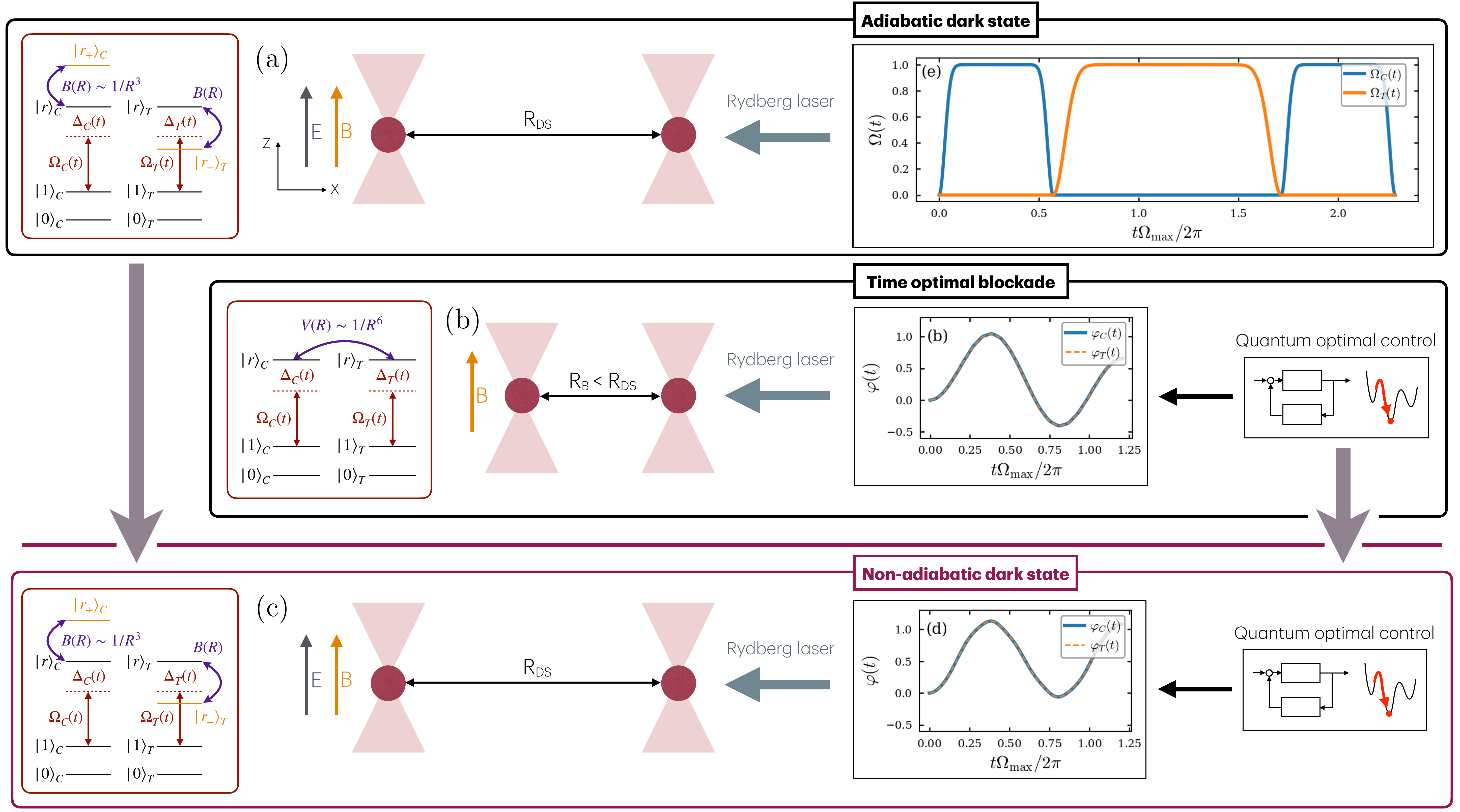}
    \caption{Concepts of blockade, adiabatic and non-adiabatic dark state gates. (a) Schematic of the adiabatic dark state gate: to realize a dark state gate experimentally, F\"orster resonance can be achieved through, for example, AC- or DC-Stark tuning, represented schematically here by an electric field applied along the quantization axis. A $\pi-2\pi-\pi$ sequence is applied to the qubits to realize the gate~\cite{petrosyan2017high}. The level diagram including the dipole-dipole coupling of Rydberg states as well as the Rabi couplings is shown in the inset. (b) Schematic of the time optimal blockade gate: the blockade gate entirely relies on the van der Waals interaction, and only a magnetic field is applied. Due to the $1/R^{6}$ dependence of the van der Waals interaction, the interatomic distance must fall below the blockade radius, putting more stringent constraint on the interatomic distance compared to the dark state gates. On the other hand, the interatomic distance for implementing the dark state gate does not suffer from this restriction (demonstrated in the figure by having $R_{\mathrm{B}} <R_{\mathrm{DS}}$). The Rabi pulses are shaped by the application of quantum optimal control (QOC) to find time optimal pulses. The optimal pulses are global~\cite{jandura2022time} with the Rabi coupling equal to the maximum allowed value, whereas the phases for the control and target qubits $\varphi_{C}(t)$ and $\varphi_{T}(t)$ are shown in the inset. (c) By combining the level scheme of dark state gates with QOC pulse shaping, a \textit{non-adiabatic dark state} gate is achieved. Similar to the blockade gate, the time optimal non-adiabatic dark state gate exhibits global pulses, with the Rabi frequencies set to their maximal value, and equal control and target pulse phases. This non-adiabatic gate design benefits from the advantages of both the adiabatic dark state gate and the time optimal blockade gate. The level scheme along with the detailed implementation of the CZ gate for each approach is explained in the text.}
    \label{fig:schematic}
\end{figure*}

One of the most widely used and experimentally mature implementations of entangling gates in neutral-atom platforms is based on the Rydberg blockade mechanism. Here we refer to such gates as \emph{blockade gates} \cite{Jaksch2000,lukin2001dipole,saffman2010quantum,shi2022quantum}. These gates exploit strong van der Waals interactions between Rydberg states to implement conditional logic operations such as controlled-phase and controlled-Z gates~\cite{saffman2010quantum,shi2022quantum}. Blockade gates offer several key advantages: gate times reduced by the application of quantum optimal control techniques~\cite{jandura2022time,ma2023high}, simple pulse sequences, and natural compatibility with large-scale neutral-atom arrays~\cite{saffman2010quantum,shi2022quantum}. A standard blockade gate can be implemented using only a small number of laser pulses and does not require cyclic evolution, auxiliary microwave fields~\cite{giudici2025fast}, or highly structured optimal-control waveforms. Despite these advantages, blockade-based gates suffer from intrinsic limitations. The dominant error sources include spontaneous emission from Rydberg states, laser phase and intensity noise, imperfect blockade due to finite interaction strengths, atomic motion, and spatial inhomogeneity, all of which lead to leakage and phase errors, as analyzed in detail in many previous works~\cite{jiang2023sensitivity,pagano2022error,giudici2025fast,shi2022quantum}. These challenges have motivated extensive efforts to improve blockade gate performance using pulse shaping~\cite{theis2016high,Levine2019}, quantum optimal control~\cite{jandura2022time,chang2023high}, adiabatic or echo-based protocols~\cite{shi2018accurate,jandura2023optimizing}, and hybrid schemes incorporating geometric phases or error-suppressing techniques~\cite{su2023rabi}. While blockade gates currently represent the leading neutral-atom entangling-gate technology, substantial room remains for improvement on the path toward fault-tolerant operation.

On the other hand, \textit{non-blockade} gate schemes offer complementary advantages to blockade-based approaches by relaxing the requirement of strong interaction-induced level shifts. In particular, dark state~\cite{petrosyan2017high,khazali2025controlled} and interference-based gates suppress population of the $\ket{rr}$ state, reducing sensitivity to atomic motion, imperfect blockade, and spatial inhomogeneity~\cite{petrosyan2017high,jin2024geometric}. Geometric and holonomic gates exploit geometric phase accumulation, providing intrinsic robustness against certain control errors such as pulse-area and timing imperfections~\cite{sun2024holonomic,jin2025nonadiabatic}. Together, these approaches provide flexible mechanisms to mitigate specific error sources. 

However, these advantages typically come at the cost of increased experimental complexity. Many non-blockade approaches rely on intricate multi-level coupling schemes, multi-photon excitation, cyclic parameter modulation, auxiliary microwave control, or finely shaped and tightly synchronized laser pulses with strict phase and amplitude requirements. Examples include holonomic and geometric gates that depend on cyclic evolution in multi-level Rydberg manifolds~\cite{sun2024holonomic}, non-adiabatic dark-path geometric gates requiring off-resonant multilevel driving~\cite{jin2024geometric}, and microwave-assisted resonant dipole-dipole gates that combine optical and microwave fields for entanglement generation~\cite{giudici2025fast,tan2018implementation}. Additionally, schemes based on Rydberg antiblockade with adiabatic passage demand finely tuned detunings and pulse sequences~\cite{tan2018implementation}. While these approaches offer additional flexibility or robustness against specific error sources, they typically come at the cost of increased experimental overhead.

In this context, an especially promising direction is to investigate whether key elements of non-blockade gate schemes can be incorporated into experimentally established blockade-based architectures with minimal modification. Achieving the robustness benefits of non-blockade protocols while retaining the simplicity, controllability, and scalability of standard blockade-gate pulse designs would represent a significant advance for neutral-atom quantum processors.

Quantum optimal control (QOC) provides a powerful framework to pursue this goal. QOC has been successfully applied across a wide range of neutral-atom and Rydberg platforms, including the preparation of exotic many-body states~\cite{cui2017optimal,mukherjee2020preparation}, generation of GHZ states~\cite{mukherjee2020bayesian}, and the realization of fast two- and three-qubit gates~\cite{jandura2022time}. In particular, time-optimal controlled-phase gates based on blockade interactions have demonstrated the ability of QOC to push gate performance close to fundamental limits~\cite{doultsinos2025fundamental}. These successes naturally motivate the question of whether optimal control can similarly be used to reformulate and improve non-blockade-based gate schemes within experimentally realistic constraints.

This leads to several key questions addressed in this work. Is it possible to realize non-adiabatic extensions of traditionally adiabatic non-blockade gates using the same pulse-shaping capabilities developed for blockade gates? What are the minimal ingredients from non-blockade schemes that must be introduced to gain their advantages, and how do these modifications compare—both conceptually and quantitatively—to established blockade-based implementations? Crucially, can such hybrid approaches achieve competitive or even superior performance relative to existing time-optimal blockade gates, while retaining experimental simplicity?

Here we demonstrate that these goals can indeed be achieved for the dark state gate. Using QOC, we construct non-adiabatic implementations of this intrinsically adiabatic gate scheme that require only smooth, experimentally feasible control pulses. The pulses obtained through QOC for the dark state gate yield a non-adiabatic realization of the gate. We refer to this non-adiabatic implementation as the \textit{non-adiabatic dark state} (NAd dark state) gate, to distinguish it from the original adiabatic dark state (Ad dark state) gate~\cite{petrosyan2017high}. Importantly, the non-adiabatic approach supports the use of global optical pulses~\cite{Levine2019,evered2023high,jandura2022time}, a feature that is highly relevant for experimental realizations of the dark state gate and represents a significant improvement over the original requirement of local addressing~\cite{petrosyan2017high}. Remarkably, the resulting gates not only achieve significantly reduced operation times, but also exhibit enhanced robustness against internal–motional coupling as well as amplitude and phase noise, particularly at larger interatomic separations. In this way, the optimized dark state gate combines the speed characteristic of time-optimal blockade gates with the intrinsic resilience of dark state-based protocols. 

A key insight underlying this improvement is the trade-off between non-adiabatic errors and gate duration. While faster, non-adiabatic evolution can increase the population leakage into doubly-excited Rydberg states, the reduced exposure time to decoherence and technical noise can compensate for this effect. We explicitly show that, within an optimal-control framework, this balance can be favorably tuned to yield a net reduction in gate error. 

Overall, our results demonstrate that non-blockade gate concepts can be meaningfully integrated into blockade-based experimental architectures without resorting to complex shortcut-to-adiabaticity techniques or substantially increasing experimental complexity. This approach enables high-fidelity, fast two-qubit gates at larger atomic separations, opening new avenues for scalable neutral-atom quantum processors that combine speed, robustness, and experimental practicality.

This work is structured as follows. Section II introduces the necessary background on the physics of the blockade and the dark state gates, and sets the notations and conventions used throughout the work. Section III discusses the pulse sequences used to implement the time optimal blockade as well as the adiabatic and non-adiabatic dark state gates. In Section IV we analyze different sources of errors for the non-adiabatic dark state gate and compare its error response to the adiabatic dark state and the blockade gate. Finally, we provide our conclusions and outlook in Section V.

\section{Two-qubit gates with Rydberg atoms}
\label{sec:rydberg_2q_gates}

Two-qubit gate implementations in neutral-atom platforms rely on encoding the computational basis states
$\{\ket{0}, \ket{1}\}$ in long-lived atomic ground or hyperfine states of alkaline and alkaline earth atoms, while entangling operations are mediated through laser-induced excitation of these states to highly excited Rydberg levels. Depending on the gate design, a set of $N_{\mathrm{Ryd}}$ Rydberg states $\ket{r_{i}}, \, i =1, \cdots, N_{\mathrm{Ryd}}$ can be involved in the gate dynamics. The strong state-dependent interactions between Rydberg excitations provide the essential nonlinearity required for entanglement generation. In the following, we introduce the formalism and notation used throughout this work to describe the gate dynamics.

For all the gate protocols considered in this work, we assume that the $\ket{1} \leftrightarrow \ket{r}$ transition can be driven via Rydberg excitation lasers for the control and target atoms independently. The Rydberg laser excitations for the control and target atoms are characterized by Rabi couplings $\Omega_{C}(t)$ and $\Omega_{T}(t)$, and phases $\varphi_{C}(t)$ and $\varphi_{T}(t)$, respectively. The Rydberg excitation Hamiltonian denoted by $\hat{H}_{\mathrm{Ryd\text{-}exc}}(t)$ takes the following form 
\begin{equation}
\label{eq:Hlaserdrive}
\begin{split}
    \hat{H}_{\mathrm{Ryd\text{-}exc}}(t)
     =
    & \Bigg(
    \frac{\Omega_{C}(t)}{2}
    e^{i\varphi_{C}(t)}
    \dyad{1}{r}_{C} + h.c. \Bigg) \\ &
    + \Bigg( \frac{\Omega_{T}(t)}{2}
    e^{i\varphi_{T}(t)}
    \dyad{1}{r}_{T}
    + \mathrm{h.c.} \Bigg) \, .
\end{split}
\end{equation}

In addition to Rydberg excitations, implementation of two-qubit gates requires interaction between the Rydberg excitations of the control and target atoms, through the dipole-dipole interaction between the two atoms described by
\begin{equation}\label{eq:Vdd}
    \hat{V}_{dd}(\vc{R}) = \frac{1}{4 \pi \epsilon_0 R^3} \Big[ \hat{\vc{d}}_{C} \cdot \hat{\vc{d}}_{T} - 3 \big( \vc{e}_{\vc{R}} \cdot \hat{\vc{d}}_{C} \big) \big ( \vc{e}_{\vc{R}} \cdot \hat{\vc{d}}_{T} \big)  \Big] \, .
\end{equation}
In Eq.~\eqref{eq:Vdd}, $\vc{R}$ is the position vector of the target atom relative to the control atom, $\vc{e}_{\vc{R}}$ is the unit vector in $\vc{R}$'s direction, and $\hat{\vc{d}}_{C}$ ($\hat{\vc{d}}_{T}$) is the dipole operator of the control (target) atom. The interaction Hamiltonian $\hat{H}_{\mathrm{Ryd}}(\vc{R})$ entering the gate dynamics is given in terms of $\hat{V}_{dd}(\vc{R})$ projected on the manifold of Rydberg states involved in the gate dynamics. Hereafter, we drop the explicit $\vc{R}$-dependence of the interaction Hamiltonian for simplicity. For each gate, the total Hamiltonian governing the gate dynamics takes the form
\begin{equation}\label{eq:HtotGate}
    \hat{H}_{\mathrm{tot}}(t) = \hat{H}_{\mathrm{Ryd\text{-}exc}}(t) + \hat{H}_{\mathrm{Ryd}} \, .
\end{equation}
We now make this construction explicit for each gate type, specifying in turn the relevant Rydberg states, the resulting interaction Hamiltonian $\hat{H}_{\mathrm{Ryd}}$, and the corresponding gate dynamics for the blockade gate and the two dark state gate protocols.

\subsection{Blockade gates}

Rydberg atoms possess large electric dipole moments and polarizabilities, leading to strong, long-range interactions at micrometer-scale separations. These gates exploit the strong van der Waals interaction between two identical Rydberg states to suppress simultaneous excitation of both atoms and thereby generate conditional dynamics. Shortly after their introduction~\cite{saffman2010quantum}, the idea to harness Rydberg blockade for performing tasks relevant for quantum computation was extended to mesoscopic ensembles, emphasizing its scalability, robustness, and applicability to quantum computing.

In the simplest realization of a blockade gate, a single Rydberg state $\ket{r}$ is involved for both atoms. The Rydberg excitation is described by the Hamiltonian in Eq.~\eqref{eq:Hlaserdrive}, while the van der Waals interaction between the control and target Rydberg excitations implements the required correlations. The total Hamiltonian of the blockade gate is therefore given by 
\begin{equation}
\label{eq:HRydBlockade}
    \hat{H}_{\mathrm{tot,Blockade}}(t) = \hat{H}_{\mathrm{Ryd\text{-}exc}}(t) + V(R) \dyad{rr}{rr} \, ,
\end{equation}
where $V(R) \!=\! C_6 / R^6$ is the van der Waals interaction potential, $R$ is the interatomic separation, and $C_6$ is the van der Waals coefficient associated with the Rydberg state $\ket{r}$, see Fig.~\ref{fig:schematic}(b).

The blockade gate is a mature entangling mechanism in neutral-atom platforms, with well-established theoretical models, benchmarks, and experimental demonstrations~\cite{Jaksch2000,lukin2001dipole,saffman2010quantum,shi2022quantum}. Gate durations can be made short compared to ground state coherence times and the rate of various kinds of errors~\cite{jandura2022time,ma2023high}, reducing sensitivity to technical noise sources and fundamental limitations of the dynamics, such as spontaneous emission or unwanted admixture with other Rydberg states~\cite{doultsinos2025fundamental}. 

In practice, the blockade interaction is always finite, leading to residual population of Rydberg excitations and leakage errors when the Rabi frequency is not negligible compared to the interaction shift. Furthermore, a well-known limitation of blockade gates arises from residual population in the Rydberg excited state $\ket{rr}$. Because the van der Waals interaction has a $1/R^6$ distance dependence, residual population in doubly-excited Rydberg states produces forces that couple internal states to atomic motion, degrading gate fidelity. This population experiences a force $F(R) \!=\! -\partial V(R) / \partial R$, which couples the internal Rydberg excitation to the relative motional degree of freedom of the atoms~\cite{petrosyan2017high}. As a consequence, atomic motion becomes entangled with the internal states, leading to motional decoherence and a reduction of gate fidelity, particularly in shallow or finite-temperature optical tweezers. The required small distance between the atoms required for high-fidelity blockade gates to be within a well-defined blockade radius, constrains the array geometry and limits interaction selectivity in dense systems.

\subsection{Non-blockade and dark state-based gates}

Beyond blockade-based schemes, several alternative proposals for two-qubit gates have been developed (e.g., Refs.~\cite{petrosyan2017high,giudici2025fast}). Among these proposals, a number of non-blockade based gates exploit geometric phases to implement entangling operations~\cite{jin2024geometric,su2023rabi,sun2024holonomic}. Many non-blockade schemes engineer entanglement through states whose energies are weakly dependent (or independent) on interatomic separation, mitigating force-induced motional entanglement that exists as an error source for the blockade gates.

Among the most promising of these non-blockade based gates is the \emph{two-atom dark state gate}~\cite{petrosyan2017high}, in which entanglement is generated through adiabatic evolution of a collective dark state. By exploiting interference, adiabatic elimination, or dark states, these gates keep the population of $\ket{rr}$ small, reducing both Rydberg decay and mechanical forces.

A key advantage of the dark state approach is that the relevant eigenstate is, to leading order, insensitive to the interatomic separation. As a result, population in the doubly-excited Rydberg manifold is strongly suppressed throughout the gate, mitigating motional decoherence and improving robustness against atomic motion.

The dark state gate involves three Rydberg states per atom, $\ket{r},\ket{r_+}$ and $\ket{r_-}$ with single-atom energies $E(r_-) \!\leq\! E(r) \!\leq\! E(r_+)$, see Fig.~\ref{fig:schematic}(a) and (c). The doubly-excited state $\ket{rr}$ is assumed to be resonantly or near-resonantly coupled to
$\ket{r_{+}r_{-}}$ and $\ket{r_{-}r_{+}}$ via dipole-dipole interaction. The dipole-dipole interaction is symmetric with respect to the exchange of the control and target atom, therefore it couples $\ket{rr}$ to the symmetric combination
\[
\ket{(r_{+}r_{-})}
=
\frac{1}{\sqrt{2}}
\Big(
\ket{r_+r_-}
+
\ket{r_-r_+}
\Big) \, ,
\]
where parentheses denote symmetric combination. 

In the absence of the dipole-dipole interaction, the energy of the initial and final pair states are related by 
\begin{equation}\label{eq:foersterCondition}
    E(r) + E(r) + \Delta
    =
    E(r_{+}) + E(r_{-}),
\end{equation}
where $\Delta$ is the Förster defect. Hereafter, we assume resonant interactions where $\Delta \!=\! 0$. This resonance condition can in principle be achieved through the application of a constant electric field to Stark-tune the resonance, see Fig.~\ref{fig:schematic}(a). In Ref.~\cite{supmat} Sec.~\ref{sec:Foerstertuning}, we consider two experimentally realistic scenarios where $\Delta \!=\! 0$ is achieved through DC- or AC-Stark tuning, and we present and compare the error rates due to imperfections in F\"orster resonance tuning in these two schemes. The gate errors due to imperfect Förster tuning depend on the specific experimental techniques used to achieve the resonance condition and on the error-mitigation strategies employed. Such error types are not universal to the gate protocols considered in this work, and we therefore do not focus on them here.

As with the blockade gate, the Rydberg excitation Hamiltonian of the dark state gate is given by $\hat{H}_{\mathrm{Ryd-exc}}(t)$ in Eq.~\eqref{eq:Hlaserdrive}. The Rydberg interaction Hamiltonian, however, corresponds to a detuned Förster resonance between the $\ket{rr}$ and the $\ket{(r_{+}r_{-})}$ states, giving a total Hamiltonian
\begin{equation}
\label{eq:HRydDarkState}
\begin{split}
    \hat{H}_{\text{tot,DarkState}}(t)
    &=\hat{H}_{\mathrm{Ryd-exc}}(t) 
    +\Delta\,
    \dyad{(r_{+}r_{-})}{(r_{+}r_{-})}
    \\
    &\quad+
    \big [
    B(R)\,
    \dyad{(r_{+}r_{-})}{rr}
    + \mathrm{h.c.}
    \big ] \, ,
\end{split}
\end{equation}
where $B(R)$ is the dipole--dipole coupling strength, scaling as $B(R) \sim R^{-3}$, see Fig.~\ref{fig:schematic}(a) and (c). 

Given the total Hamiltonian governing the dynamics of a two-atom system, the question of gate design with neutral atoms amounts to finding Rydberg excitation pulses that implement the unitary operation corresponding to the desired gate. In what follows, we explain this procedure for the blockade gate as well as the adiabatic and non-adiabatic versions of the dark state gate.

\section{Design of pulse sequences}

A gate design problem reduces to finding a suitable level scheme and pulse sequences which drive coherent state transfer between the considered states such that at the end of the pulse sequence, the desired unitary operator corresponding to the gate operation is implemented. Here we discuss two prominent methods to obtain pulse sequences for the implementation of a CZ gate.

A figure of merit to analyze the performance of a two-qubit gate is the trace fidelity
\begin{equation}\label{eq:fidelityTrace}
    F = \bigg | \frac{1}{d} \mathrm{Tr}\big( \hat{U}_{\mathrm{target}}^{\dagger} \hat{U}_{\mathrm{tot}}(T) \big) \bigg |^2 \, ,
\end{equation}
where $\hat{U}_{\mathrm{target}}$ stands for the target unitary, $\hat{U}_{\mathrm{tot}}(t) \!=\! \mathcal{T} \big\{ \mathrm{exp} \big( - i/\hbar \int^{t}_{0} dt' \, \hat{H}_{\mathrm{tot}}(t') \big)\big\}$ is the implemented unitary, which is the time evolution operator generated by the total Hamiltonian $\hat{H}_{\mathrm{tot}}(t)$ of the corresponding gate, and $d \!=\! 2^2$ is the dimension of the two-qubit gate's computational subspace. The fidelity error for the gate under study is then defined by $\varepsilon \!=\! 1-F$.

The target unitary for a CZ gate takes the form $\hat{U}_{\mathrm{target}} \!=\!\mathrm{diag}\big(1,e^{i\theta_1},e^{i\theta_2},e^{i(\theta_1+\theta_2+\pi)}\big)$ in the computational basis $\{\ket{00},\ket{01},\ket{10},\ket{11}\}$. Here, $\theta_1$ and $\theta_2$ are arbitrary single-qubit phases gained by the qubit state $\ket{1}$ relative to $\ket{0}$, and the gate is diagonal in the computational basis. Here we discuss two methods for obtaining such pulses. First, we describe the \jak sequence, which was used in the original proposal~\cite{Jaksch2000} and is considered in the main work introducing the dark state gate~\cite{petrosyan2017high}. Then, we apply QOC to find optimal pulses to implement the non-adiabatic dark state CZ gate.

\subsection{\jak sequence}
\label{sec:jakseq}

The $\pi-2\pi-\pi$ pulse was the first pulse shape proposed to implement the CZ gate. It works by (i) applying a $\pi$ pulse to bring the state $\ket{11}$ to $-i\ket{r1}$, (ii) applying a $2\pi$ pulse on the target qubit to bring the state $-i\ket{r1}$ to $-i\ket{r1}$ (since the Rabi oscillation is suppressed due to the Rydberg blockade), and (iii) applying a final $\pi$ pulse on the control qubit to bring $-i\ket{r1}$ to $-\ket{11}$. It can be shown that this pulse sequence maps $\ket{00} \to \ket{00}$, $\ket{01} \to -\ket{01}$ and $\ket{10} \to -\ket{10}$, thus it implements a CZ phase up to a $\pi$ single qubit phase of states $\ket{1}$. The adiabatic version of this protocol was considered in Ref.~\cite{petrosyan2017high} to implement an adiabatic dark state gate. Here, instead of the strong Rydberg blockade, the evolution of the adiabatic state over the zero energy dark state leads to a vanishing phase accrual at step (ii), thus implementing the same gate.

Here we consider smooth Rabi coupling pulse shapes for the target and control qubits, of the form
\begin{equation}\label{eq:pi2pipipulses}
    \Omega(t) = \Omega_{\mathrm{max}} \bigg[ b(T/\sigma) - a(T/\sigma) \Big( e^{-t^2/2\sigma^2} + e^{-(t-T)^2/2\sigma^2}\Big)\bigg] \, ,
\end{equation}
where $a(x) \!=\! 1/\big( 1+e^{-x^2/2}+e^{-x^2/8} \big)$, $b(x) \!=\! \big(1+e^{-x^2/2}\big)/\big( 1+e^{-x^2/2}+e^{-x^2/8} \big)$, $\Omega_{\mathrm{max}}$ is the maximum Rabi frequency, $T$ is the duration that implements a pulse with pulse area $\Theta$ and is obtained from 
\begin{equation}
    T \Omega_{\mathrm{max}} = \frac{\Theta}{b(T/\sigma) - \sqrt{2\pi} \, (\sigma/T) \, a(T/\sigma) \mathrm{erf} \big( T/\sqrt{2} \sigma \big)} \, ,
\end{equation}
with $\mathrm{erf}(x) \!=\! 2/\sqrt{\pi}\int^{x}_{0} dt \, e^{-t^2/2}$. The ratio $\sigma/T$ controls how smoothly the Rabi pulse reaches the maximum value $\Omega_{\mathrm{max}}$, which we set to $\sigma/T \!=\! 0.05$. These pulse forms are used in the original proposal~\cite{petrosyan2017high} and are depicted in the inset of Fig.~\ref{fig:schematic}(a). 

In the original implementation of the dark state gate, the Rabi pulses are adiabatic, such that the instantaneous eigenstates in step (ii) follows adiabatically the two-atom dark state 
\begin{equation}\label{eq:Dt}
    \ket{D(t)} \!=\! 1/\sqrt{B^2 + \Omega^2_{T}(t)/4} \, \big[ B \ket{r1} - \Omega_{T}(t)/2 \ket{(r_{+}r_{-})} \big] \, .
\end{equation}
In this way, the population of the $\ket{rr}$ state is suppressed and the force due to the van der Waals interaction is mitigated. However, non-adiabaticity can result in a finite population of $\ket{rr}$, which affects the gate performance. 

\begin{figure}
    \centering
    \includegraphics[width=0.48\textwidth]{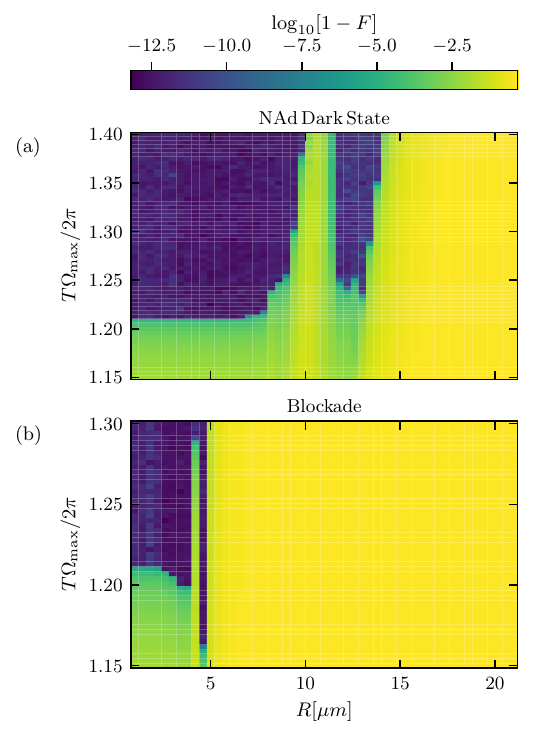}
    \caption{Fidelity error for the CZ gate as a function of interatomic distance $R$ and implementation time $T$ (in units of $2\pi/\Omega_{\mathrm{max}}$) for (a) the non-adiabatic dark state gate, and (b) for the blockade gate. The resonant Rydberg states are $\ket{r} \!=\! \ket{70 P_{3/2},m_j \!=\! 3/2}$, $\ket{r_{+}} \!=\! \ket{71 S_{1/2},m_j \!=\! 1/2}$ and $\ket{r_{-}} \!=\! \ket{70 S_{1/2},m_j \!=\! 1/2}$. In the $R \!\to\! 0$ limit, the minimum implementation time approaches the optimal time $T_{\mathrm{opt}} \simeq 1.2113 \, \cdot 2\pi/\Omega_{\mathrm{max}}$ for the ideal blockade gate ($\Omega_{\mathrm{max}} = 2\pi \times 10 \, \mathrm{MHz}$ ), i. e. the blockade gate with $V \to \infty$. As the blockade radius $R_{B} \simeq 4.6 \, \mu m$ is approached, the blockade gate rapidly looses performance, while the non-adiabatic dark state gate still maintains high fidelity for implementation times $T$ comparable to $T_{\mathrm{opt}}$ at distances $R \sim 10 \, \mu m$.}
    \label{fig:vdWB_NADS_comparison}
\end{figure}

\begin{figure*}
    \centering
    \includegraphics[width=\textwidth]{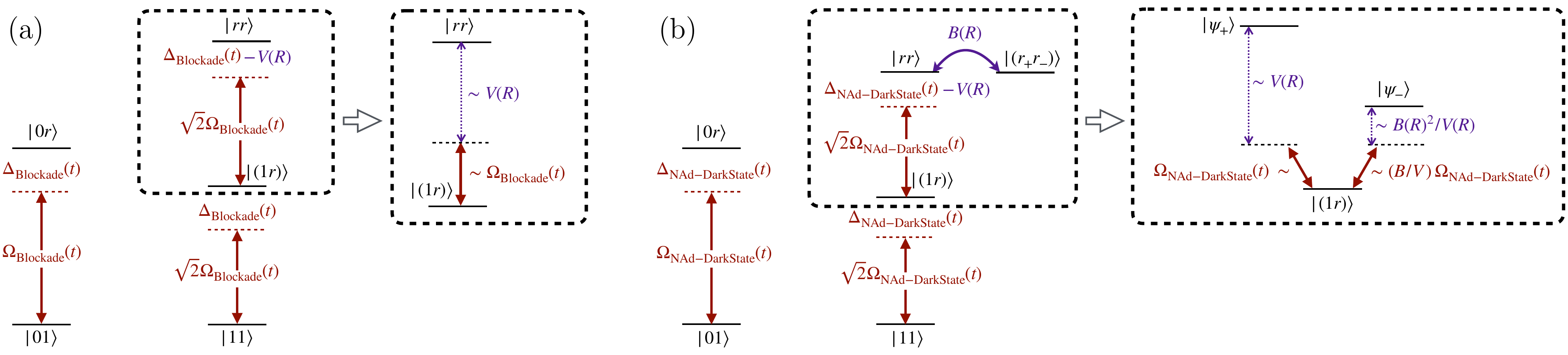}
    \caption{Level diagrams illustrating the two-atom states, energies, and couplings underlying the dynamics of (a) the blockade gate and (b) the non-adiabatic dark state gate. In both schemes, $\ket{01}$ ($\ket{10}$) is Rabi-coupled to the singly-excited Rydberg state $\ket{0r}$ ($\ket{r0}$), and $\ket{11}$ is coupled to the symmetric state $\ket{(1r)} \!=\! (\ket{1r}+\ket{r1})/\sqrt{2}$ with the same Rabi frequency. (a) In the blockade gate, $\ket{(1r)}$ is further coupled to the doubly-excited state $\ket{rr}$ with strength $\sqrt{2}\,\Omega_{\mathrm{Blockade}}(t)$, while $\ket{rr}$ is shifted by the van der Waals interaction $V(R)$. As $R\to 0$, $V(R)\to\infty$, detuning $\ket{rr}$ far from resonance with $\ket{(1r)}$ and effectively removing it from the dynamics. (b) In the non-adiabatic dark state gate, $\ket{rr}$ is coupled to the doubly-excited state $\ket{(r_{+}r_{-})}$ with strength $B(R)$, hybridizing the two into dressed states $\ket{\psi_{\pm}}$. The state $\ket{\psi_{+}}$ (predominantly $\ket{rr}$) is detuned from $\ket{(1r)}$ by $\sim V(R)$ and retains a coupling to $\ket{(1r)}$ of order $\Omega_{\mathrm{NAd\text{-}DarkState}}(t)$, while $\ket{\psi_{-}}$ (predominantly $\ket{(r_{+}r_{-})}$) is detuned from $\ket{(1r)}$ by $\sim B(R)^{2}/V(R) \to C_{3}^{2}/C_{6}$ and couples to $\ket{(1r)}$ only weakly, with strength $\sim \big( B(R)/V(R) \big)\,\Omega_{\mathrm{NAd\text{-}DarkState}}(t) \to 0$ as $R\to 0$. Consequently, in the $R\to 0$ limit, an effective two-level system formed by $\ket{11}$ and $\ket{(1r)}$ emerges in the dynamics of both gates, decoupled from all doubly-excited states~\cite{Levine2019,fromonteil2023protocols,jandura2022time}.}
    \label{fig:blockDynamics}
\end{figure*}

\subsection{Application of QOC: the non-adiabatic dark state CZ gate}
\label{sec:AppQOC}

Quantum optimal control has previously been applied to the design of time optimal entangling gates, most notably in Ref.~\cite{jandura2022time}, where high fidelity time optimal blockade CZ gates were realized. In contrast, its application to dark state gate schemes has not been explored prior to the present work. Here, we apply QOC using the well-established gradient-based algorithm GRadient Ascent Pulse Engineering (GRAPE). To find pulses that implement high-fidelity gates, the control signals consist of the Rabi couplings of the control and target qubits $\Omega_{C,T}(t)$, and their phases $\varphi_{C,T}(t)$. Fixing an implementation time $T$, time is then discretized into $N$ equal length intervals $\delta t  \!=\! T/N$ with discrete points $t_{k} \!=\! k \delta t,\, k \!=\! 0,1,2, \cdots, N$ with $t_0 \!=\! 0$ and $t_{N} \!=\! T$. The time evolution unitary is then implemented as $\hat{U}_{\mathrm{tot}}(T) \!=\! \prod^{N-1}_{k = 0} \, \hat{U}_{\mathrm{tot}}(t_{k+1},t_{k})$ where $\hat{U}_{\mathrm{tot}}(t_{k+1},t_{k}) \!=\! \mathcal{T} \big\{ \mathrm{exp} \big( -i/\hbar \int^{t_{k+1}}_{t_k} d\tau \, \hat{H}_{\mathrm{tot}}(\tau) \big) \big\}$. In the $[t_{k},t_{k+1})$ interval, the control signals are assumed to be constant $\Omega_{C,T}(t) \!=\! \Omega_{C,T}(t_k) \!\equiv\! \Omega_{C,T}[k]$, $\varphi_{C,T}(t) \!=\! \varphi_{C,T}(t_k) \!\equiv\! \varphi_{C,T}[k]$, $t \!\in\! [t_{k},t_{k+1})$, and the Hamiltonian in the $k$'th time interval becomes only a function of these control signals. The cost function for the optimization is the fidelity error $\varepsilon$ which is a function of all the control signals, $\varepsilon[\{\Omega_{C,T}[k]\}^{N}_{k=0},\{\varphi_{C,T}[k]\}^{N}_{k=0}]$, which can be minimized using a gradient-based optimization algorithm, where the L-BFGS-B algorithm is often used. Here we use the \texttt{Python} implementation of L-BFGS-B using the function \texttt{scipy.optimize.minimize} To obtain time optimal pulses, the fixed gate time $T$ is reduced and optimal pulses which implement high-fidelity gates are obtained until a sharp increase in error is found.

We apply QOC using GRAPE to find time-optimal pulses for both the blockade and dark state gates, yielding the non-adiabatic dark state gate in the latter case. To benchmark the performance of each gate, we consider an experimentally relevant setup consisting of $^{133}$Cs atoms, with the Rydberg state $\ket{r} \!=\! \ket{n P_{3/2},m_j \!=\! 3/2}$, which can be tuned to a Förster resonance with states $\ket{r_{-}} \!=\! \ket{n S_{1/2},m_j \!=\! 1/2}$ and $\ket{r_{+}} \!=\! \ket{(n+1) S_{1/2},m_j \!=\! 1/2}$. We use the \texttt{PairInteraction} package~\cite{weber2017calculation} to find the interaction strengths, decay rates of the states, and other relevant system parameters. We assume a magnetic field $B_{z} \!=\! 100\,\mathrm{G}$, with the quantization axis taken along $z$ and the interatomic axis oriented orthogonally, along $x$. This large magnetic field Zeeman-shifts the sublevels of the $P$-state manifold far off resonance, thereby separating these leakage states from the target Rydberg state $\ket{nP_{3/2},m_j \!=\! 3/2}$. For instance, using the \texttt{PairInteraction} package, we find the Zeeman splitting between $\ket{70P_{3/2},m_j \!=\! 3/2}$ and $\ket{70P_{3/2},m_j \!=\! 1/2}$ to be approximately $2\pi \times 202\,\mathrm{MHz}$. Here, we take the $n \!=\! 70$ Rydberg state of $^{133}$Cs, which has a blockade radius $R_{B} \simeq 4.6 \, \mu m$ at $\Omega_{\mathrm{max}} \!=\! 2\pi \times 10 \, \mathrm{MHz}$.  

Applying quantum optimal control to both the blockade gate and the dark state gate, we find that at the minimal gate implementation time, $\Omega_{C}(t) \!=\! \Omega_{T}(t) \!=\! \Omega_{\mathrm{max}}$, and $\varphi_{C}(t) \!=\! \varphi_{T}(t)$, see Fig.~\ref{fig:schematic}(b,c). This observation is consistent with the earlier findings~\cite{jandura2022time} that the time optimal blockde gate is realized for global pulses with constant Rabi frequencies equal to the maximum achievable Rabi frequency. The infidelities for different gate times and distances for the two atoms are shown in Fig.~\ref{fig:vdWB_NADS_comparison}. 

We find that at small distances within the blockade radius, the non-adiabatic dark state gate performs similar to the blockade gate. Specifically, the minimum implementation time for both gates approach $T \, \Omega_{\mathrm{max}}/2\pi \simeq 1.2113$, which is the value obtained for the time optimal blockade gate in the limit $V\!\to\! \infty$ in Ref.~\cite{jandura2022time}. The existence of such a limiting value reflects the fact that the optimal control solution obeys a minimum-action principle, as encapsulated by Pontryagin’s maximum principle. This behavior has been explicitly demonstrated for the time-optimal ideal blockade gate in Ref.~\cite{jandura2022time}. 

To understand the behavior of the gate fidelities as $R\!\to\!0$, a natural approach is to compare the relative strengths of the Rydberg interaction terms and the Rabi couplings appearing in $\hat{H}_{\mathrm{tot}}(t)$ for the blockade and non-adiabatic dark state gates. For both the blockade gate~\cite{jandura2022time} and the dark state gate~\cite{petrosyan2017high}, the time evolution of the $\ket{01}$, $\ket{10}$ and $\ket{11}$ states proceeds via independent sets of level couplings, as illustrated in Fig.~\ref{fig:blockDynamics}. Since the global drive scheme used in both gates renders the coupling $\ket{10}\leftrightarrow\ket{r0}$ identical to $\ket{01}\leftrightarrow\ket{0r}$, only the latter is shown explicitly in Fig.~\ref{fig:blockDynamics}.

In the blockade gate, the van der Waals shift $V(R)$ diverges as $R\to 0$. Thus, the $\ket{rr}$ state is detuned far from $\ket{(1r)}$ and is effectively removed from the dynamics. The time evolution of $\ket{11}$ is then well described by a single effective two-level system, $\{\ket{11},\ket{(1r)}\}$, coupled only via the single-excitation Rydberg transitions, see Fig.~\ref{fig:blockDynamics}(a).

In the non-adiabatic dark state gate, the corresponding interaction is the resonant dipole-dipole coupling $B(R)\sim C_{3}/R^{3}$ between $\ket{rr}$ and $\ket{(r_{+}r_{-})}$, which hybridizes these states into two dressed eigenstates, denoted by $\ket{\psi_{\pm}}$. Since $V(R)$ diverges faster than $B(R)$ as $R\to 0$ ($R^{-6}$ versus $R^{-3}$), the ratio $B(R)^{2}/V(R)$, which sets the detuning of $\ket{\psi_{-}}$ from $\ket{(1r)}$, approaches a finite constant $\sim C_{3}^{2}/C_{6}$, while the residual coupling between $\ket{\psi_{-}}$ and $\ket{(1r)}$, scaling as $B(R)/V(R)$, vanishes. The state $\ket{\psi_{-}}$ therefore decouples from the dynamics in this limit. The state $\ket{\psi_{+}}$, in turn, plays the same role that $\ket{rr}$ plays in the blockade gate: its detuning from $\ket{(1r)}$ is of order $V(R)$, while its Rabi coupling to $\ket{(1r)}$ remains of order $\Omega_{\mathrm{max}}$. Since $V(R)\to\infty$ as $R\to 0$, $\ket{\psi_{+}}$ likewise decouples from the dynamics in this limit, see Fig.~\ref{fig:blockDynamics}. As a result, the evolution of $\ket{11}$ is governed effectively by the dynamics of the two-level system $\{\ket{11},\ket{(1r)}\}$, in close analogy with the blockade gate. Since the dynamics of both gates reduce to the same set of effective two-level systems as $R\!\to\! 0$, quantum optimal control (QOC) applied to either scheme must converge to identical optimal pulse shapes and identical optimal fidelities in this limit. 

As the interatomic distance approaches the blockade radius $R_{B} \simeq 4.6\,\mu\mathrm{m}$, the blockade gate rapidly loses fidelity, whereas the dark state gate maintains high fidelity while its implementation time remains close to optimal, $T \simeq T_{\mathrm{opt}}$. This behavior reflects the different scaling of the two relevant interactions with distance: the van der Waals interaction falls off rapidly as $R^{-6}$, while the resonant dipole-dipole interaction decreases more slowly, as $R^{-3}$ (see Fig.~\ref{fig:schematic}). We note that although Ref.~\cite{jandura2022time} demonstrates that time-optimal gates remain effective beyond the strong-blockade regime, their weakest-blockade case ($V/\Omega_{\mathrm{max}} \simeq 10$, see Sec.~6 of Ref.~\cite{jandura2022time}) still corresponds to an intermediate coupling regime, within the blockade radius. By contrast, our analysis extends beyond the blockade radius, where $V/\Omega_{\mathrm{max}} \simeq 1$, into a regime of substantially weaker interaction strength ($V/\Omega_{\mathrm{max}}$ reaching $1.46 \times 10^{-4}$ at $R\!=\! 20 \, \mu m$).

These results demonstrate that the non-adiabatic dark state protocol obtained via QOC retains the intrinsic robustness of adiabatic dark state gates against reduced interaction strengths and increased interatomic spacing, while simultaneously overcoming their known limitation of long gate times. In this way, the optimized non-adiabatic dark state gate combines the favorable distance scaling and robustness of dark state schemes with gate speeds comparable to those of time optimal blockade gates, enabling high fidelity operation beyond the conventional blockade radius.

Having established the superior performance of the non-adiabatic dark state gate at longer distances compared to the blockade gate, an important question remains: how susceptible is the non-adiabatic dark state gate to the coupling of the Rydberg excitation to atomic motion compared to its adiabatic counterpart and the blockade gate? This question is critical since the non-adiabatic protocol increases the contribution of the $\ket{rr}$ state in the gate dynamics, thus it is expected that the resulting van der Waals force would be more prominent and more strongly affect the gate performance. In the next section, we analyze the response of the two variations of the dark state gate to atomic motion and compare it to the blockade gate.

\vspace{-0.2cm}

\section{Comparative error analysis for the non-adiabatic dark state gate}

In the reminder of this work, we focus on the comparison of the non-adiabatic gate performance to its adiabatic counterpart and the blockade gate in terms of other prominent error sources, namely the decoherence due to the coupling of Rydberg excitations to the atomic motion, the decay of Rydberg states, and the influence of the phase and amplitude noise.

\subsection{Influence of the internal-motional coupling on the gate performance}

The coupling of internal degrees of freedom with atomic motion is a source of error in the implementation of quantum gates with Rydberg atoms~\cite{giudici2025fast,jandura2023optimizing,evered2023high,tsai2025benchmarking}. The effect of internal-motional coupling on the dynamics of the Rydberg systems have been studied in detail for applications in quantum simulation and many-body physics~\cite{zhang2024motional,ates2007many,magoni2023molecular,mazza2020vibrational,gambetta2020engineering,parvej2025lamb}. Thus, understanding the dynamics of these systems and the influence of this coupling in the performance of the quantum engineered systems and devising methods to mitigate the resulting errors is of key interest in these fields. 

In the context of two-qubit gates, the quantum fluctuations of the atomic positions inside the tweezer traps couple to Rydberg excitations via the resulting induced changes in the Rydberg interaction strength, as well as sampling the electric field of the driving laser at different positions. The former effect is responsible for causing an effective force between the two atoms~\cite{petrosyan2017high}, while the latter results in a position-fluctuation induced random phase of the Rydberg transition, which can also be described in terms of photon recoil \cite{giudici2025fast}.

Here we focus on a configuration similar to Ref.~\cite{giudici2025fast} where the excitation laser propagates along the interatomic axis, which we take to be the $x$ axis. For every phonon number state of the motional degree of freedom of each atom within the trap, $\ket{n_a},\,n_a\!=\!0,1,\cdots$, the infidelity is calculated by time-evolving the total internal-motional state according to the total Hamiltonian in Ref.~\cite{giudici2025fast} and then averaging over the thermal phonon distribution at a certain temperature $T_{\mathrm{th,trap}}$ and trap frequency $\omega_{x}$. The resulting infidelities due to the coupling to the thermal motion are calculated for $T_{\mathrm{th,trap}} \!=\! 1 \, \mu K$ and trap frequencies $\omega_{x} \!=\! 50 \, \mathrm{kHz}$. The principal quantum number of the Rydberg excitation is taken to be $n\!=\!70$, and the gates are compared at a distance $R \simeq 4.2 \, \mu m$, where all gates demonstrate high fidelity ($\varepsilon < 10^{-10}$). Surprisingly, the non-adiabatic gate shows the least fidelity error, $\varepsilon_{\mathrm{NAd-DarkState}} \!\simeq \! 1.11 \times 10^{-5}$, which is smaller than the adiabatic counterpart with $\varepsilon_{\mathrm{Ad-DarkState}} \!\simeq \! 4.00 \times 10^{-5}$. This counterintuitive effect can be attributed to the competition between the unwanted population of $\ket{rr}$ state due to non-adiabaticity, and the total gate duration time, which increases the error rate. Here, while the non-adiabatic dynamics increases the population of the undesired Rydberg state, the shorter gate duration time leads to a smaller error compared to the adiabatic gate which implements the gate more slowly. Furthermore, both types of the dark state gate demonstrate higher fidelities as compared to the blockade gate, although the implementation of the blockade gate possesses time optimality. The dark state gates are expected to exhibit reduced sensitivity to fluctuations in atomic motion, since the leading-order correction to the gate dynamics in the limit $B \!\to\! \infty$ scales as $(\Omega_{\mathrm{max}}/B)^2$. To provide intuition for this scaling, in Ref.~\cite{supmat} Sec.~\ref{supmatsec:ExactSolution} we derive the exact eigenstates and eigenenergies governing the dark state dynamics connected to $\ket{11}$, for general time-dependent Rabi frequencies $\Omega_{C}(t)$ and $\Omega_{T}(t)$, in the limit where the van der Waals interaction is negligible and perfect Förster resonance is achieved. All in all, the lower fidelity error of the non-adiabatic dark state gate highlights its potential, and more broadly the promise of dark state gates for achieving better performance compared to the blockade gates by application of quantum optimal control methods.

In this section we only considered the influence of the atomic motion parallel to the laser excitation propagation direction, as in Ref.~\cite{giudici2025fast}. This assumption is justified because the atomic displacement perpendicular to the laser excitation is also perpendicular to the two-atom relative position. Thus, the leading order contribution of the transverse displacement comes only in the second order in the atomic distance fluctuations. This contribution is orders of magnitude smaller than the contribution considered here, and therefore negligible.

\subsection{Errors due to the Rydberg decay}
\label{sec:DecayError}

Spontaneous emission of Rydberg states together with their vulnerability to black-body radiation are known to be major error sources in quantum hardware based on neutral atoms. Here we examine the fidelity of the three types of gates under study with respect to the decay of Rydberg states. For all gate types considered in this work, the gate dynamics incorporating the decay of the relevant Rydberg states is governed by the non-hermitian Hamiltonian $\hat{H}_{\mathrm{Decay}}(t) = \hat{H}_{\mathrm{tot}}(t) + \hat{H}_{\mathrm{NH}} $, where
\begin{equation}\label{eq:HamiltonianDecay}
    \hat{H}_{\mathrm{NH}} = - \! \sum^{N_{\mathrm{Ryd}}}_{j=1}\,\frac{i\hbar \Gamma_{j}}{2} \bigg( \hat{\Pi}_{j,C}  + \hat{\Pi}_{j,T} \bigg) \, ,
\end{equation}
accounts for the decay of the Rydberg states. In Eq.~\eqref{eq:HamiltonianDecay}, $\Gamma_{j}$ is the decay rate of the Rydberg state $\ket{r_j}$, while $\hat{\Pi}_{j,C}$ ($\hat{\Pi}_{j,T}$) is the projector onto the state $\ket{r_j}$ of the control (target) atom. 

For each gate type, the effect of Rydberg state decay on the gate fidelity is captured by the non-Hermitian evolution under $\hat{H}_{\mathrm{Decay}}(t)$. This is justified since the Rydberg decay rates $\Gamma_j$ are slow compared to the gate time $T$, i.e., $\Gamma_j T \ll 1$, placing the system in the Fermi golden rule regime. The non-unitary evolution operator $\hat{U}_{\mathrm{Decay}}(t)$ generated by $\hat{H}_{\mathrm{Decay}}(t)$ yields a decay error $\varepsilon_{\mathrm{Decay}} \!=\! 1 - F_{\mathrm{Decay}}$, where
\begin{equation}
    F_{\mathrm{Decay}} = \left|\frac{1}{d}\,\mathrm{Tr}\!\left(\hat{U}^{\dagger}_{\mathrm{target}}\, \hat{U}_{\mathrm{Decay}}(T)\right)\right|^2.
\end{equation}
Rydberg state decay introduces an additive correction $\delta\varepsilon_{\mathrm{Decay}}$ to the gate infidelity, so that $\varepsilon_{\mathrm{Decay}} \!=\! \varepsilon + \delta\varepsilon_{\mathrm{Decay}}$.

To evaluate $\delta\varepsilon_{\mathrm{Decay}}$, the key quantities are the instantaneous populations of the single-excitation states $\ket{r_j q}$, with one atom in a Rydberg state $r_j$ and the other in a qubit state $q \!\in\! \{0,1\}$, and the double-excitation states $\ket{r_j r_k}$, with both atoms in Rydberg states. For each initial state $\ket{p}$, $p \!\in\! \{01,10,11\}$, these populations are defined under the unperturbed evolution $\hat{U}_{\mathrm{tot}}(t)$ as
\begin{equation}\label{eq:Pqjp}
    P_{jq|p}(t) = \bra{p} \hat{U}^{\dagger}_{\mathrm{tot}}(t) \dyad{r_j q}{r_j q} \hat{U}_{\mathrm{tot}}(t)\ket{p} \, ,
\end{equation}
and similarly $P_{jk|p}(t) \!=\! \bra{p} \hat{U}^{\dagger}_{\mathrm{tot}}(t) \dyad{r_j r_k}{r_j r_k} \hat{U}_{\mathrm{tot}}(t)\ket{p}$.

\begin{table}[h]
    \centering
    \begin{tabular}{| c || c | c | c |}
        \hline
        \multicolumn{4}{|c|}{$T_{\mathrm{th}} \!=\! 298 \, \mathrm{K}$} \\
        \hline
        $n$ & $\Gamma_{nP_{3/2}}$ (kHz) & $\Gamma_{nS_{1/2}}$ (kHz) & $\Gamma_{(n+1)S_{1/2}}$ (kHz) \\
        \hline
        40 & 17.2 & 30.9 & 28.9 \\
        \hline
        50 & 9.5  & 16.4 & 15.6 \\
        \hline
        60 & 6.0  & 10.0 & 9.6  \\
        \hline
        70 & 4.1  & 6.7  & 6.5  \\
        \hline\hline
        \multicolumn{4}{|c|}{$T_{\mathrm{th}} \!=\! 4 \, \mathrm{K}$} \\
        \hline
        $n$ & $\Gamma_{nP_{3/2}}$ (kHz) & $\Gamma_{nS_{1/2}}$ (kHz) & $\Gamma_{(n+1)S_{1/2}}$ (kHz) \\
        \hline
        40 & 6.9 & 17.6 & 16.2 \\
        \hline
        50 & 3.2 & 8.4 & 7.8 \\
        \hline
        60 & 1.8 & 4.6 & 4.3 \\
        \hline
        70 & 1.1 & 2.8 & 2.7 \\
        \hline
    \end{tabular}
    \caption{
    Decay rates of the Rydberg states involved in the gate dynamics
    at two different temperatures.
    }
    \label{tab:parameters}
\end{table}

From the conditional instantaneous populations $P_{jq|p}(t)$ and $P_{jk|p}(t)$, we define the conditional instantaneous loss rates
\begin{equation}
\kappa_{jq|p}(t) = \frac{\Gamma_j}{2} P_{jq|p}(t), \qquad \kappa_{jk|p}(t) = \frac{\Gamma_{j} + \Gamma_{k}}{4} P_{jk|p}(t),
\end{equation}
due to Rydberg decay from the states $\ket{r_j q}$ and $\ket{r_j r_k}$, respectively. The corresponding conditional cumulative losses are
\begin{equation}\label{eq:Lambdas}
\Lambda_{jq|p}(t) \equiv \int^{t}_0 dt' \,\kappa_{jq|p}(t'), \qquad \Lambda_{jk|p}(t) \equiv \int^{t}_0 dt' \,\kappa_{jk|p}(t').
\end{equation}
\begin{figure*}
    \centering
    \includegraphics[width=1\textwidth]{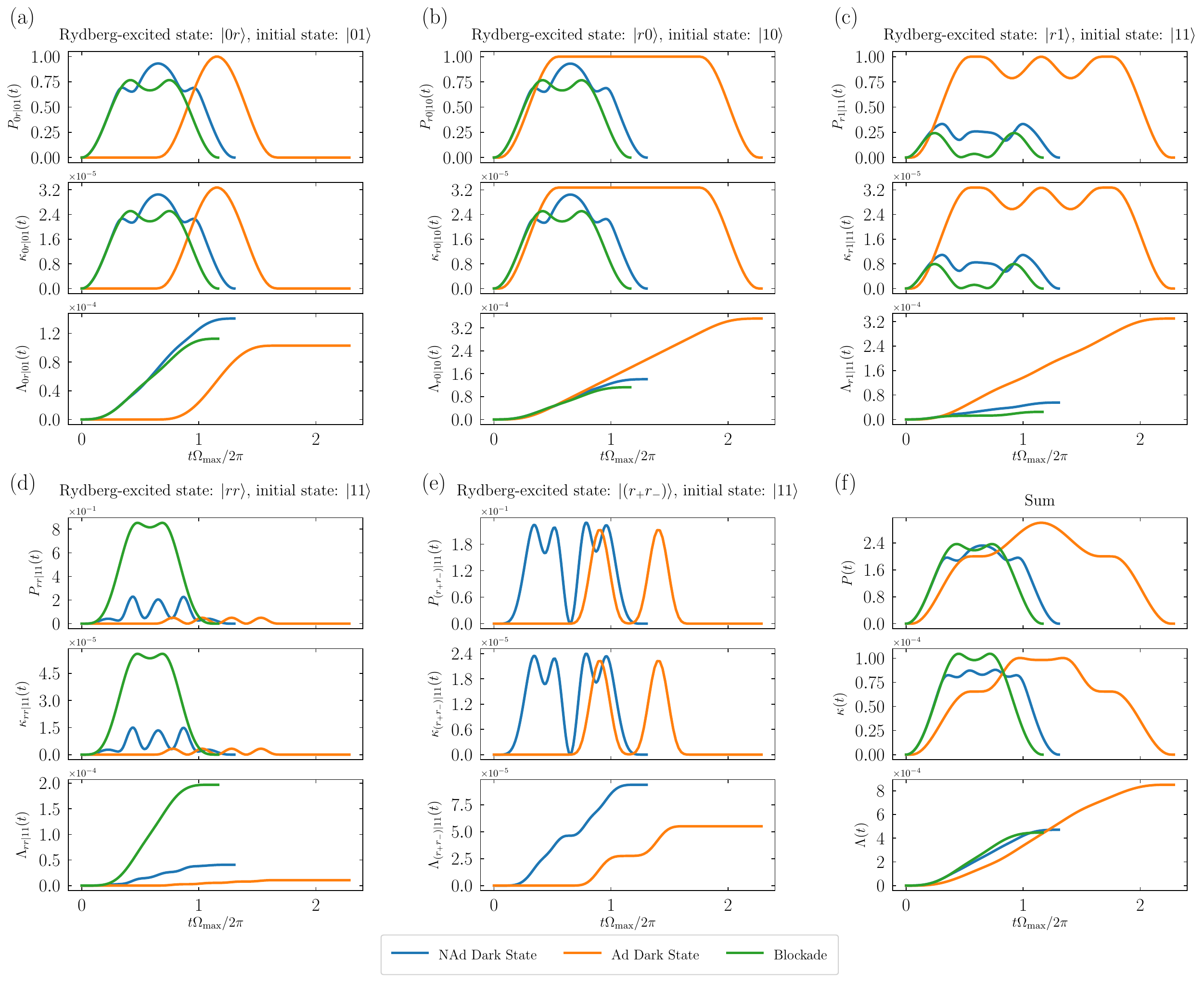}
    \caption{Population of the intermediate Rydberg-excited states, $P$, instantaneous loss rate, $\kappa$, and cumulative loss, $\Lambda$, conditioned on the initial state, for the non-adiabatic dark state gate (blue), adiabatic dark state gate (orange), and blockade gate (green), as a function of time, for the transitions (a) $\ket{01}\leftrightarrow\ket{0r}$, (b) $\ket{10}\leftrightarrow\ket{r0}$, (c) $\ket{11}\leftrightarrow\ket{r1}$, (d) $\ket{11}\leftrightarrow\ket{rr}$, and (e) $\ket{11}\leftrightarrow\ket{(r_+ r_-)}$. Panel (f) shows the corresponding sums, $P(t) = \sum_{p}\big(\sum_{q}\sum_{j}P_{jq|p}+\sum_{j,k}P_{jk|p}\big)$, $\kappa(t) = \sum_{p}\big(\sum_{q}\sum_{j}\kappa_{jq|p}+\sum_{j,k}\kappa_{jk|p}\big)$, and $\Lambda(t) = \sum_{p}\big(\sum_{q}\sum_{j}\Lambda_{jq|p}+\sum_{j,k}\Lambda_{jk|p}\big)$, with $p\in\{01,10,11\}$, $q\in\{0,1\}$, and $j,k = 1,\dots,N_{\mathrm{Ryd}}$. In (a), despite its longer gate time, the adiabatic gate accumulates a smaller integrated population than the non-adiabatic and blockade gates, giving it a smaller $\Lambda_{0r|01}$ at the end of the protocol. Because the blockade and non-adiabatic gates are global, $\Lambda_{r0|10}$ follows a similar trend to $\Lambda_{0r|01}$. The adiabatic gate, however, requires Rydberg excitation of the control qubit during step (ii) of its protocol, resulting in a substantially larger $\Lambda_{r0|10}$ than for the other two gates --- a pattern that persists for the $\ket{11}\leftrightarrow\ket{r1}$ transition in (c). For the doubly-excited state $\ket{rr}$ in (d), the blockade gate yields a markedly larger final $\Lambda_{rr|11}$ than either dark state protocol, reflecting the effectiveness of population suppression in the dark state approach. Nevertheless, non-adiabatic effects leave $\Lambda_{rr|11}$ larger for the non-adiabatic than the adiabatic gate. The state $\ket{(r_+r_-)}$ in (e) is not populated during the blockade-gate dynamics but contributes to the decay error of both dark state gates, with $\Lambda_{(r_+r_-)|11}$ again larger for the non-adiabatic implementation due to non-adiabatic effects. Panel (f) shows that, although the adiabatic gate exhibits considerably lower cumulative loss for the doubly-excited transitions and for $\ket{01}\leftrightarrow\ket{0r}$, its markedly larger losses on the $\ket{10}\leftrightarrow\ket{r0}$ and $\ket{11}\leftrightarrow\ket{r1}$ transitions dominate the total, yielding the largest overall cumulative loss among the three protocols. Comparing the non-adiabatic and blockade gates, the former exhibits larger cumulative loss on the $\ket{01}\leftrightarrow\ket{0r}$, $\ket{10}\leftrightarrow\ket{r0}$, and $\ket{11}\leftrightarrow\ket{r1}$ transitions, and carries an additional contribution from $\ket{11}\leftrightarrow\ket{(r_+r_-)}$. However, its substantially lower loss on the $\ket{11}\leftrightarrow\ket{rr}$ transition brings its total cumulative loss close to that of the blockade gate, and significantly below that of the adiabatic gate.}
    \label{fig:detaileddecayanalysis}
\end{figure*}
The decay error $\delta\varepsilon_{\mathrm{Decay}}$ is then expressed, to first order in $\Gamma_{j}T$, directly in terms of these cumulative losses evaluated at the gate time $T$
\begin{equation}\label{eq:dgsimple}
\begin{split}
    \delta \varepsilon_{\mathrm{Decay}} & \simeq 
     \sum_{q =0,1 } \! \sum^{N_{\mathrm{Ryd}}}_{j=1} \frac{2}{d}\sum_{p\in\{01,10,11\}} \Big[  \Lambda_{jq|p}(T) + (j \leftrightarrow q) \Big]  
    \\
    & + \sum^{N_{\mathrm{Ryd}}}_{j,k=1} \frac{2}{d} \sum_{p\in\{01,10,11\}} \Big[  \Lambda_{jk|p}(T) + (j \leftrightarrow k) \Big] \, ,
\end{split}
\end{equation}
to the first order in $\Gamma_j T$.  A derivation of Eq.~\eqref{eq:dgsimple} is provided in Ref.~\cite{supmat} Sec.~\ref{supmatsec:DecayInfidelity}.

\begin{figure}
    \centering
    \includegraphics[width=0.48\textwidth]{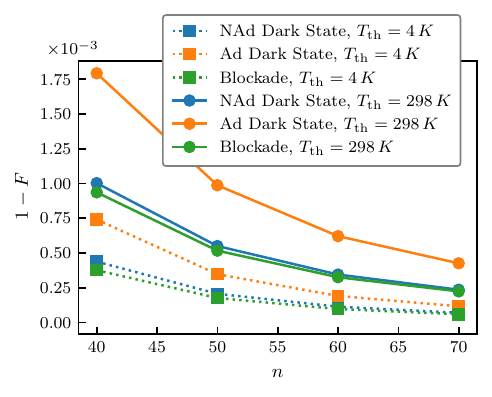}
    \caption{Gate error due to Rydberg-state decay as a function of the principal quantum number $n$, for the non-adiabatic dark state (blue), adiabatic dark state (orange), and blockade (green) gates, at temperatures $T_{\mathrm{th}} \!=\! 4\,\mathrm{K}$ and $T_{\mathrm{th}} \!=\! 298\,\mathrm{K}$, corresponding to cryostat and room-temperature experimental conditions, respectively. The non-adiabatic dark state gate shows markedly improved performance relative to its adiabatic counterpart, closely following the blockade gate.}
    \label{fig:decay}
\end{figure}

The time dependence of the instantaneous populations, instantaneous decay rates, and cumulative losses for the initial states $\ket{01}$, $\ket{10}$, and $\ket{11}$ is shown in Fig.~\ref{fig:detaileddecayanalysis}. The Rydberg state decay rates used to evaluate the instantaneous decay rates and cumulative losses are obtained from the \texttt{PairInteraction} package for the relevant Rydberg states, including black-body radiation at temperatures $T_{\mathrm{th}} \!=\! 4\,\mathrm{K}$ and $T_{\mathrm{th}} \!=\! 298\,\mathrm{K}$, corresponding to cryostat and room-temperature experiments, respectively. The resulting decay rates are tabulated in Table~\ref{tab:parameters}. We show only the two-atom Rydberg-excited states whose populations are non-vanishing to within machine precision. For all gate protocols considered, these correspond to the transitions $\ket{01} \leftrightarrow \ket{0r}$, $\ket{10} \leftrightarrow \ket{r0}$, and $\ket{11} \leftrightarrow \ket{r1}$ for two-atom states with a single Rydberg excitation, and to $\ket{11} \leftrightarrow \ket{rr}$ and $\ket{11} \leftrightarrow \ket{(r_{+}r_{-})}$ for two-atom states with two Rydberg excitations. We note that the dynamics of both the non-adiabatic gate and the blockade gate couple $\ket{11} \leftrightarrow \ket{1r}$. However, since the protocols are global, the corresponding population and decay quantities are identical to those of the $\ket{11} \leftrightarrow \ket{r1}$ transition. For the adiabatic dark state gate, by contrast, the $\ket{11} \leftrightarrow \ket{1r}$ transition vanishes identically, and we therefore omit it from Fig.~\ref{fig:detaileddecayanalysis}.

The population of $\ket{0r}$ shown in Fig.~\ref{fig:detaileddecayanalysis}(a) is smaller for the adiabatic gate than for the non-adiabatic and blockade gates, with the non-adiabatic dark state gate incurring the largest loss. However, because the adiabatic dark state gate relies on a perfect transfer of the control qubit from $\ket{1}$ to $\ket{r}$ to implement the intermediate step (ii) described in Sec.~\ref{sec:jakseq}, the cumulative loss from the $\ket{r0}$ channel is significantly larger for the adiabatic protocol than for the non-adiabatic and blockade gates. Interestingly, the peak loss rate associated with this control-qubit excitation is comparable across all three gates (see $\kappa_{r0|10}(t)$ and $\Lambda_{r0|10}(t)$ in Fig.~\ref{fig:detaileddecayanalysis}(b)). However, since the gate durations of the non-adiabatic and blockade protocols are roughly half that of the adiabatic dark state gate, their final decay errors in this channel remain smaller. Similarly, Rydberg excitation of the control atom drives the population of $\ket{11}$ almost entirely into $\ket{r1}$, with a small fraction transferred to $\ket{(r_+r_-)}$ owing to its coherent admixture with $\ket{r1}$ in the two-atom dark state (panels (c) and (e) of Fig.~\ref{fig:detaileddecayanalysis}, respectively). By contrast, the population of $\ket{r1}$ in the non-adiabatic and blockade gates is considerably smaller than in the adiabatic gate, which, combined with their shorter duration, yields a much smaller cumulative loss from Rydberg decay in this channel. As a result, the overall decay error of the adiabatic gate is considerably larger than that of the other two protocols due to decay from intermediate states containing a single Rydberg excitation.

In contrast, the suppressed population of doubly-excited two-atom states $\ket{rr}$ in the adiabatic protocol, relative to the other two gates, results in a much smaller decay contribution due to population in $\ket{rr}$. Notably, the corresponding increase in cumulative loss for the non-adiabatic gate is only marginal relative to the adiabatic gate, indicating that it largely inherits the favorable suppression of its adiabatic counterpart in this channel, unlike the blockade gate. The blockade gate, meanwhile, exhibits no decay from $\ket{(r_+r_-)}$, whereas both dark state gates incur a finite loss from this channel due to transient population of $\ket{(r_+r_-)}$. This loss is larger for the non-adiabatic gate than for the adiabatic gate, as expected from its non-adiabaticity.

Summing the losses from the singly- and doubly-excited states for each gate, Fig.~\ref{fig:detaileddecayanalysis}(f) shows that the adiabatic gate has the largest total error among the three protocols: its larger decay from the $\ket{r0}$ and $\ket{r1}$ channels, combined with its longer gate time, outweighs the benefit of its suppressed decay from the doubly-excited states. The non-adiabatic and blockade gates, in contrast, exhibit closely comparable population dynamics across the intermediate excited states, giving the non-adiabatic protocol a total error only slightly higher than the blockade gate. This small difference arises from a trade-off: while the non-adiabatic gate has a significantly lower cumulative loss from $\ket{rr}$ than the blockade gate, its slightly higher losses from the $\ket{0r}$, $\ket{r0}$, $\ket{r1}$, and $\ket{(r_+r_-)}$ states ultimately outweigh this advantage, yielding a marginally higher overall decay error than the blockade gate.

Finally, the resulting decay errors are shown in Fig.~\ref{fig:decay} for all three gates as a function of the principal quantum number $n$. Across all values of $n$ and both temperatures considered, the non-adiabatic dark state gate significantly outperforms its adiabatic counterpart. Moreover, its decay error closely tracks that of the blockade gate, indicating that the non-adiabatic protocol achieves performance comparable to the blockade gate.

\subsection{Fidelity error due to the laser frequency noise}

Aside from the aforementioned error sources, fluctuations in the laser frequency noise is known to be another major source of error. The frequency fluctuations directly translates into the Rabi phase fluctuations. Thus, studying the influence of the phase fluctuation is also of prime importance to quantify the gate performance.

\begin{figure}
    \centering
    \includegraphics[width=0.5\textwidth]{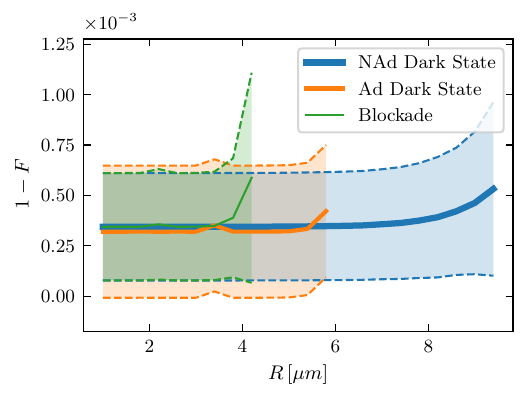}
    \caption{Mean infidelity due to laser phase fluctuations (thick solid lines) for the non-adiabatic dark state gate (blue) and the blockade gate (orange). The regions within one standard deviation is highlighted for each gate. While at distances $R \lesssim 3 \, \mu m$ all gates have similar fidelity errors, the blockade gate rapidly loses fidelity around $R \simeq 4 \, \mu m$, while the non-adiabatic dark state gate maintains a reasonable performance up to $R \simeq 8 \, \mu m$.}
    \label{fig:phasenoise}
\end{figure}

To this end, $\mathcal{N}$ random phase time traces $\delta \varphi^{(i)}_{a}(t), i \!=\! 1,\cdots , \mathcal{N}$ is created for atom $a$ ($a\!\in\!\{C,T\}$) according to the single-side-band power spectral density of the laser phase noise, see Sec.~\ref{supmatsec:PhaseNoiseAnalysis} of Ref.~\cite{supmat}. The noise samples are then added to the pulse obtained by quantum optimal control for qubit $a$, $\varphi_{a}(t)$, to obtain noisy phase pulse time traces $\varphi^{(i)}_{a}(t) \!=\! \varphi_{a}(t) + \delta \varphi^{(i)}_{a}(t)$. The obtained infidelities $\varepsilon^{(i)}_{\mathrm{phase}} \!=\! \varepsilon \big[ \{\Omega_{a}(t),\varphi^{(i)}_{a}(t) \}_{a=C,T} \big]$ form a sequence of random phase errors with mean $\bar{\varepsilon}_{\mathrm{phase}} \!=\! 1/\mathcal{N} \, \sum^{\mathcal{N}}_{i=1} \varepsilon^{(i)}_{\mathrm{phase}}$ and standard deviation $\sigma^2_{\mathrm{phase}} \!=\! \sum^{\mathcal{N}}_{i=1} \big(\varepsilon^{(i)}_{\mathrm{phase}} - \bar{\varepsilon}_{\mathrm{phase}} \big)^2/(\mathcal{N}-1)$. The mean error and its variance are depicted in Fig.~\ref{fig:phasenoise} for the three type of gates considered in this work for various distances. It is observed that in the $R \!\to\! 0$ limit, all gates reach the ideal blockade noise response. 

As the blockade radius is approached, the blockade gate shows increasing susceptibility to noise, whereby both the mean fidelity error and its variance increase. On the other hand, the adiabatic dark state gate exhibits greater robustness to variations in the interatomic distance, while its performance gradually degrades with increasing separation and eventually drops sharply at the largest distance considered ($R_{B}\!\lesssim\! R\simeq6 \,\mu m$ in Fig.~\ref{fig:phasenoise} in the atomic distance). The non-adiabatic gate demonstrates the best robustness against the noise: while in the blockade regime, the fidelity error statistics closely follows that of the blockade gate, it sustains over distances $R \gtrsim  8\, \mu m$, where eventually the non-adiabatic dark state gate starts to deteriorate. This result also demonstrates the enhanced performance in terms of phase noise resilience of the non-adiabatic gate compared to both the blockade and the adiabatic dark state gate.

\vspace{-0.3cm}

\subsection{Fidelity error due to the laser intensity noise}

The last error source we consider is the shot-to-shot laser intensity fluctuations. The effect of this error source is modeled by obtaining the infidelity as a function of the change in Rabi frequency $\Omega_{a}$ by $\delta \Omega_{a}$ ($a\in\{C,T\}$) which, due to its low frequency can be taken as constant during the application of the gate \cite{fromonteil2023protocols,jandura2023optimizing}.

\begin{figure}
    \centering
    \includegraphics[width=0.44\textwidth]{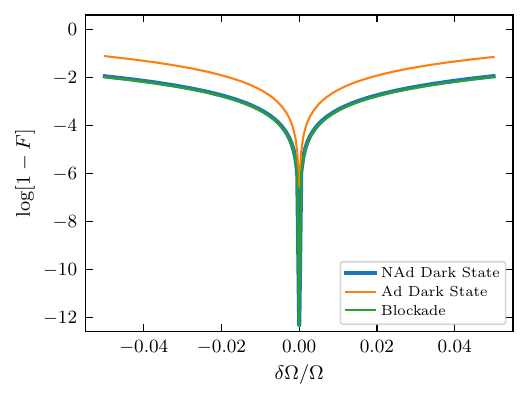}
    \caption{Infidelity due to the shot-to-shot intensity noise, for the non-adiabatic dark state (blue), adiabatic dark state (orange), and the blockade gate (green). The performance of the non-adiabatic dark state gate is enhanced compared to the adiabatic dark state gate and becomes comparable to the blockade gate.}
    \label{fig:intensity}
\end{figure}

The resulting error for different gates are depicted in Fig.~\ref{fig:intensity}, where it can be inferred that the non-adiabatic dark state gate performs similarly to the blockade gate due to the intensity fluctuations, while both gates are more robust compared to the adiabatic dark state gate. The adiabatic dark state gate is known to be sensitive to the fluctuations in the Rabi frequency, as it causes a remaining population in the Rydberg states after the application of the gate \cite{petrosyan2017high}. We conclude that the performance of the dark state gate in response to intensity fluctuations can be enhanced by application of quantum optimal control to make the gate non-adiabatic.

\section{Conclusion and outlook}

We applied quantum optimal control to design high-fidelity, time-optimal pulses for the Rydberg dark state gate, originally introduced as an adiabatic protocol~\cite{petrosyan2017high}, and performed a systematic, quantitative comparison with standard blockade-based gate implementations. The resulting non-adiabatic dark state gate retains the robustness to variations in interatomic distance characteristic of the dark state mechanism and the long-range resonant dipole-dipole interaction, while outperforming the adiabatic gate in response to Rydberg state decay and laser amplitude and phase fluctuations. Its performance in decay error approaches that of the blockade gate, while it matches or exceeds the blockade gate for noise and distance robustness. These results demonstrate, through a unified and systematic analysis, that quantum optimal control can systematically enhance the speed, fidelity, and resilience of dark state gates.

As an outlook, it would be interesting to examine in more detail the interplay between implementation time, non-adiabaticity, and Rydberg state population, and how this competition influences gate fidelity. Another important direction is to develop control schemes that render the gates robust against Förster defect fluctuations, internal-motional coupling, the residual van der Waal interaction of the $\ket{rr}$ state and laser crosstalk for the case of different control and target Rydberg states, extending the applicability of dark state gates in realistic conditions. Designing robust pulses for longer interatomic distances could mitigate errors arising from motional states, potentially increasing the gate range while preserving high fidelity. Furthermore, exploring the effects of different excitation directions on reducing the effect of atomic motion will further illuminate the limitations and opportunities for optimizing gate performance.

\section{Acknowledgment}

This work is funded by the German Federal Ministry of Education and Research within the funding program “Quantum Technologies - from basic research to market” under Contract No. 13N16138. R.M. acknowledges support from the U.S. National Institute of Standards and Technology (NIST) through the CIPP program under Award No. 60NANB24D218. and the U.S. National Science Foundation (NSF) through the NSF TIP program under Award No. 2534232.

\bibliography{references}


\renewcommand{\thesection}{S\arabic{section}}
\renewcommand{\theequation}{S\arabic{equation}}
\renewcommand{\thefigure}{S\arabic{figure}}
\renewcommand{\thetable}{S\arabic{table}}

\onecolumngrid


\setcounter{section}{0}
\setcounter{equation}{0}
\setcounter{figure}{0}
\setcounter{table}{0}

\clearpage

\section*{SUPPLEMENTAL MATERIAL\\ ``High-fidelity non-adiabatic dark state gates for neutral atoms''}
\setcounter{page}{1}
\begin{center}
Nader Mostaan$^{1}$, Kapil Goswami$^{2}$, Peter Schmelcher$^{1,2}$, Rick Mukherjee$^{1,3,4}$\\
\emph{\small $^1$Zentrum für Optische Quantentechnologien, Universität Hamburg, Luruper Chaussee 149, 22761 Hamburg, Germany}\\
\emph{\small $^2$The Hamburg Centre for Ultrafast Imaging, Universität Hamburg, Luruper Chaussee 149, 22761 Hamburg, Germany}\\
\emph{\small $^3$Department of Physics $\&$ Astronomy, University of Tennessee, Chattanooga, TN 37403, USA}\\
\emph{\small $^4$UTC Quantum Center, University of Tennessee, Chattanooga, TN 37403, USA}\\
\end{center}

\section{Phase noise analysis}
\label{supmatsec:PhaseNoiseAnalysis}
Here we follow Refs.~\cite{jiang2023sensitivity} and ~\cite{deLeseleuc2018} to analyze the performance of the adiabatic and non-adiabatic dark state gates as well as the time optimal blockade gate under realistic laser phase noise. To this end, it is more convenient to first model the noise of the laser frequency $2\pi \nu(t) \!=\! d\varphi(t)/dt$. The laser frequency noise can be modeled as a wide-sense stationary ergodic random process, with time independent expected value $\mathbb{E}[\nu(t)] \!=\! 0$ and autocorrelation $R_{\nu\nu}(\tau)=\mathbb{E}[\nu(t) \nu(t+\tau)]$, where $\mathbb{E}[X]$ stands for the expectation value of a random variable $X$. According to the Wiener-Khinchin theorem, the autocorrelation function is the inverse Fourier transform of the power spectral density $S_{\nu}(f)$,

\begin{equation}\label{eq:PSDdef}
    R_{\nu \nu}(\tau) = \int df \, e^{-i2\pi f \tau} \, S_{\nu}(f) \, .
\end{equation}
Several noise sources contribute to the laser frequency noise power spectral density $S_{\nu}(f)$. One of the main components of $S_{\nu}(f)$ comes from the fundamental linewidth of the laser, first evaluated by Schawlow and Tones~\cite{schawlow1958infrared} and later refined \cite{lax1967classical,stephan2005laser} to be a white noise given by the power spectral density
\begin{equation}
    S^{\mathrm{QNL}}_{\nu}(f) = \frac{h f_{\mathrm{cav}} \big( \delta f_{\mathrm{cav}} \big)^2}{\bar{P}} \, ,
\end{equation}
where $f_{\mathrm{cav}} \!=\! c/\lambda_{L}$ is the laser frequency, $\delta f_{\mathrm{cav}}$ is the FWHM laser cavity linewidth and $\bar{P}$ is the time-averaged laser power. 

The next component contributing to $S_{\nu}(f)$ is the quantum noise caused by the laser medium, whose frequency response can be modeled by a second order low-pass filter with a transfer function $H_{\mathrm{LP,2nd}}(s)$ given by
\begin{equation}\label{eq:HLP2nd}
    H_{\mathrm{LP,2nd}}(s)=\frac{f_{rlx}^2}{s^2 + 2 \gamma f_{rlx} s + f^2_{rlx}} \, .
\end{equation}
In Eq.~\eqref{eq:HLP2nd}, $H_{\mathrm{LP,2nd}}(s)$ is the medium transfer function, which is the Laplace transform with the complex variable $s$ of the time-domain impulse response function of the medium $h_{\mathrm{LP,2nd}}(t)$, i.e. $H_{\mathrm{LP,2nd}}(s) \!=\! \int^{\infty}_{0} dt \, e^{-st} \, h_{\mathrm{LP,2nd}}(t)$~\cite{siebert1986circuits}, $f_{rlx}$ is the relaxation oscillation frequency, and $\gamma$ is the damping factor~\cite{svelto1998principles}. Furthermore, due to the intensity-dependent changes in the refractive index, the fundamental noise level is enhanced by a factor $\alpha^2$, with $\alpha$ the linewidth enhancement factor~\cite{henry2003theory}. The resulting quantum noise contribution to $S_{\nu}(f)$ then becomes
\begin{equation}
    S^{\mathrm{QN}}_{\nu}(f) = \alpha^2 |H_{\mathrm{LP,2nd}}(if)|^2 \, S^{\mathrm{QNL}}_{\nu}(f)
\end{equation}
It is also known that $S_{\nu}(f)$ includes a low frequency component which scales proportional to $1/f$, $S^{1/f}_{\nu} \!=\! a_{\nu}/f$~\cite{stephan2005laser,schiemangk2014accurate,mercer19911}, with $a_{\nu}$ called the flicker frequency noise coefficient. Adding all the contributions, the frequency noise power spectral density takes the following form,
\begin{equation}\label{eq:snu}
    S_{\nu}(f) = S^{1/f}_{\nu}(f) + S_{\nu}^{\mathrm{QNL}}(f) + S^{\mathrm{QN}}_{\nu}(f) =  \frac{a_{\nu}}{f} + \frac{h f_{\mathrm{cav}} \big( \delta f_{\mathrm{cav}} \big)^2}{\bar{P}} \bigg[ 1+ \alpha^2 \frac{f^{4}_{rlx}}{(f^2_{rlx} - f^2)^2 + 4 \gamma^2 f^{2}_{rlx} f^2}\bigg] \, .
\end{equation}
In the experiments, the laser frequency is stabilized via closed-loop electronic feedback (also known as the servo loop), which suppressed the low frequency noise levels~\cite{freihe,petermann1991laser}. The closed-loop electronic feedback transfer function can be modeled by
\begin{equation}
    H_{L}(s) = \sqrt{a_s} + (1-\sqrt{a_s}) H_{\mathrm{HP},1}(s) \, \Big( 1+ (\sqrt{a_p}-1) H_{\mathrm{BP}}(s) \Big) \, ,
\end{equation}

\begin{table}[h]
    \centering
    \begin{tabular}{| c | c | c | c | c | c | c | c | c | c | c |}
        \hline
        $a_{\nu}$ (MHz) & $\lambda_{L}$ (nm) & $\delta f_{\mathrm{cav}}$ (GHz) & $\bar{P}$ (mW) & $\alpha$ & $f_{rlx}$ (GHz) & $\gamma$ & $a_{s}$ & $a_{p}$ & $f_{c}$ (MHz) & $\gamma_{p}$\\
        \hline
        10 & 318.6 & 4 & 500 & 5 & 10 & 0.125 & 0.1 & 2 & 1 & 1 \\
        \hline
    \end{tabular}
    \caption{
    Laser and stabilizing servo system parameters used to obtain the phase noise power spectral density and the phase noise time series.
    }
    \label{tab:laserparams}
\end{table}
where $a_s <1$ is the residual noise floor suppression factor, $H_{\mathrm{HP,1}}(s) \!=\!s/(s+f_{c})$ is a high-pass filter which recovers the high frequency noise power spectral density above the servo bandwidth $f_c$, $H_{\mathrm{BP}}(s) \!=\! 2\gamma_{p} f_p s /\big  (s^2 + 2 \gamma_p f_p s + f_p^2 \big )$ is a band-pass filter modeling the resonant response of the feedback loop near its unit gain frequency $f_p$, $a_p$ is the servo bump amplitude, representing the excess noise factor near $f_p$, and the servo damping $\gamma_p$ controls the servo peak width. A realistic model of the effect of the servo loop can be achieved by using the empirical formula $f_{p} \!=\! f_{c}(1+2\gamma_{p})\sqrt{a_{p}}$. The filter parameters as well as the noise parameters are given in Table.~\ref{tab:laserparams}. The output frequency noise power spectral density $S_{\nu,\mathrm{out}}(f)$ is then related to the input noise $S_{\nu,\mathrm{in}}(f)$ via
\begin{equation}\label{eq:Snuout}
    S_{\nu,\mathrm{out}}(f) = |H_{L}(if)|^2 \, \big[ S_{\nu,\mathrm{in}}(f) - S_{\nu,\mathrm{QNL}}(f) \big] + S_{\nu,\mathrm{QNL}}(f) \, ,
\end{equation}
where $S_{\nu,\mathrm{in}}(f)$ is given by Eq.~\eqref{eq:snu}. Equation.~\eqref{eq:Snuout} ensures that the output noise does not fall below the fundamental noise limit. The phase noise power spectral density is then obtained from $S_{\nu,\mathrm{out}}(f)$ in Eq.~\eqref{eq:Snuout} by $S_{\phi}(f) \!=\! S_{\nu,\mathrm{out}}(f)/f^2$.

Having set the form of $S_{\phi}(f)$, we now consider the problem of phase noise time series generation from $S_{\phi}(f)$. Using the definition of the auto-correlation function and the Wiener-Khinchin theorem, for an ergodic process the power spectral density can be obtained by
\begin{equation}\label{eq:PSDtimeDer}
    S_{x}(f) = \lim_{T_{\mathrm{meas}} \to \infty} \frac{1}{T_{\mathrm{meas}}} \Big| \mathcal{F} \big\{ x_{T_{\mathrm{meas}}}(t) \big\}(f) \Big|^2 \, ,
\end{equation}
where $x(t)$ is a realization of a general random process $X(t)$, $\mathcal{F}\big\{ x(t) \big\}(f) \!=\! \int^{\infty}_{-\infty} dt \, e^{i2\pi ft} \, x(t)$ is the Fourier transform of the signal $x(t)$, and $x_{T_{\mathrm{meas}}}(t)$ is the truncation of the signal $x(t)$ over the time interval $[-T_{\mathrm{meas}}/2,T_{\mathrm{meas}}/2]$,
\begin{equation}
    x_{T_{\mathrm{meas}}}(t) = 
    \begin{cases}
        x(t), & t \leq |T_{\mathrm{meas}}/2|, \\
        0, & \text{otherwise}.
    \end{cases}
\end{equation}
We assume that the time interval $[-T_{\mathrm{meas}}/2,T_{\mathrm{meas}}/2]$ is discretized into $2N_s$ intervals, with $t_i \!=\! i \,\Delta t$ the time bins, $\Delta t \!=\! T_{\mathrm{meas}}/2N_s$ and $i \!=\! -N_s,\cdots,N_s$. The discretized power spectral density then takes the form
\begin{equation}\label{eq:SxD}
    S_{x}(f_k) = \frac{1}{T_{\mathrm{meas}}} \Bigg| \sum^{N_s}_{i=-N_s} \Delta t \, e^{i2\pi f_k t_i} \, x(t_i)\Bigg|^2 =\frac{\Delta t^2}{T_{\mathrm{meas}}} |\hat{x}(f_k)|^2 \, ,
\end{equation}
with $\hat{x}(f_k) \!=\! \sum^{N_s}_{i=-N_s} e^{i2\pi f_k t_i} \, x(t_i)$ the discrete Fourier transform of $x(t_i)$ at frequencies $f_{k} \!=\! k /T_{\mathrm{meas}}, \, k=1,\cdots,N_{s}$ and $f_0 \!=\! 0$. Note that for real signals, $\mathcal{F} \{x(t)\}(f)^{*} \!=\! \mathcal{F} \{x(t)\}(-f)$, which means $S_{x}(-f) \!=\! S_{x}(f)$. Therefore it is sufficient to consider only the positive frequencies $f>0$ and work with the single-side-band power spectral density $S_{x,\mathrm{SSB}}(f) \!=\! 2S_{x}(f)$. From Eq.~\eqref{eq:SxD}, we can obtain the amplitude of the frequency series $\hat{x}(f_k)$,
\begin{equation}\label{eq:xhatf}
    |\hat{x}(f_k)| = \sqrt{ \frac{T_{\mathrm{meas}}}{2 \Delta t^2} S_{x,\mathrm{SSB}}(f_k) } \, ,
\end{equation}
While Eq.~\eqref{eq:xhatf} is enough for producing time series for an ergodic process, it gives more accuracy if we further treat Eq.~\eqref{eq:SxD} as relating $S_{x}(f_k)$ to the ensemble average of $|\hat{x}(f_k)|^2$ over many realizations of the random process. Assuming a normal distribution for $|\hat{x}(f_k)|$, we then set 
\begin{equation}
    S_{x}(f_k) = \frac{\Delta t^2}{T_{\mathrm{meas}}} \sigma_{x}(f_k)^2 \, ,
\end{equation}
where $\sigma_{x}(f_k)^2 \!=\! \langle |\hat{x}(f_k)|^2 \rangle$ is the variance of the normally distributed random variable $|\hat{x}(f_k)|$. We now assume $x(t)$ is the phase noise signal, i.e. $x(t) \!=\! \varphi(t)$. To construct phase noise samples we use the formula
\begin{equation}
    \sigma_{\varphi} (f_k) = \sqrt{ \frac{T_{\mathrm{meas}}}{2 \Delta t ^2} S_{\varphi,\mathrm{SSB}}(f_k)} \, ,
\end{equation}
where $T_{\mathrm{meas}} \!=\! 5 \, ms$ is the total measurement time of the phase noise, $\Delta t \!=\! T_{\mathrm{meas}} /(2N_s+1)$ with $N_s \!=\! 5 \times 10^{6}$ the number of noise sample bins, $\sigma^{2}_{\varphi} (f_k)$ is the variance of a normally distributed random variable representing the magnitude of the phase noise at frequency $f_{k} \!=\! k /T_{\mathrm{meas}}, \, k=1,\cdots,N_{s}$. The time domain random samples of the phase noise are obtained by taking the discrete Fourier transform of the frequency domain samples generated according to this normal distribution for the amplitude, and a uniform distribution for the phase of the samples.

\section{Evaluation of the gate infidelities due to Rydberg state decay}
\label{supmatsec:DecayInfidelity}

In this section, we derive the expression for $\delta \varepsilon_{\mathrm{Decay}}$ in terms of the conditional cumulative losses, introduced in Sec.~\ref{sec:DecayError} of the main text. To evaluate the decay error, it is convenient to formulate the decay dynamics generated by the non-Hermitian Hamiltonian $\hat{H}_{\mathrm{NH}}$ (Eq.~\eqref{eq:HamiltonianDecay} of the main text) in the interaction picture. The non-unitary part of the time evolution in the interaction picture, $\hat{U}_{\mathrm{Decay,I}}(t)$, is defined through $\hat{U}_{\mathrm{Decay,I}}(t) \!=\! \hat{U}_{\mathrm{tot}}^{\dagger}(t) \, \hat{U}_{\mathrm{Decay}}(t)$, where $\hat{U}_{\mathrm{tot}}(t)$ and $\hat{U}_{\mathrm{Decay}}(t)$ are generated by $\hat{H}_{\mathrm{tot}}(t)$ and $\hat{H}_{\mathrm{Decay}}(t) \!=\! \hat{H}_{\mathrm{tot}}(t)+\hat{H}_{\mathrm{NH}}$, respectively. Writing the interaction-picture Hamiltonian as $\hat{H}_{\mathrm{NH,I}}(t) \!=\! \hat{U}_{\mathrm{tot}}^{\dagger}(t) \hat{H}_{\mathrm{NH}} \hat{U}_{\mathrm{tot}}(t)$, we have $\hat{U}_{\mathrm{Decay,I}}(t) \!=\! \mathcal{T} \big\{ \mathrm{exp} \big( - i/\hbar \int^{t}_{0} dt' \, \hat{H}_{\mathrm{NH,I}}(t') \big)\big\}$. The fidelity $F_{\mathrm{Decay}}$ can then be written as $F_{\mathrm{Decay}} \!=\! |g + \delta g_{\mathrm{Decay}}|^2$, where $g\!=\! (1/d)\, \mathrm{Tr} \Big( \hat{U}^{\dagger}_{\mathrm{target}} \hat{U}_{\mathrm{tot}}(T) \Big) $ and

\begin{equation}\label{eq:deltagdecay}
    \delta g_{\mathrm{Decay}} = (1/d)\, \mathrm{Tr} \Big( \hat{U}^{\dagger}_{\mathrm{target}} \hat{U}_{\mathrm{tot}}(T) \big(\hat{U}_{\mathrm{Decay,I}}(T) - \mathbb{I} \big) \Big) \, .
\end{equation}
In the weak-decay regime, $\hat{H}_{\mathrm{NH,I}}(t)$ can be treated perturbatively. To first order in the decay rates $\Gamma_j$, the expansion reads

\begin{equation}\label{eq:UDecayO1}
    \hat{U}_{\mathrm{Decay,I}}(T) - \mathbb{I} \simeq -\frac{i}{\hbar} \int^{T}_0 dt \, \hat{H}_{\mathrm{NH,I}}(t) \, .
\end{equation}
The expression for $\delta g_{\mathrm{Decay}}$ simplifies further if we set $\hat{U}^{\dagger}_{\mathrm{target}} \hat{U}_{\mathrm{tot}}(T) \!\simeq\! \mathbb{I}$, which is justified because the ideal dynamics generated by $\hat{H}_{\mathrm{tot}}(t)$ correspond to a high-fidelity gate. This approximation is equivalent to setting $g \simeq 1$, so that

\begin{equation}\label{eq:deltagDecayU}
    \delta g_{\mathrm{Decay}} \simeq (1/d)\, \mathrm{Tr} \Big( \hat{U}_{\mathrm{Decay,I}}(T) - \mathbb{I} \Big) \, ,
\end{equation}
and the decay error is given by $\delta \varepsilon_{\mathrm{Decay}} \!=\! 1-F_{\mathrm{Decay}} \! \simeq \! -2 \delta g_{\mathrm{Decay}}$.

To proceed, we expand $\hat{\Pi}_{j,C} + \hat{\Pi}_{j,T}$ appearing in the definition of $\hat{H}_{\mathrm{NH}}$ in Eq.~\eqref{eq:HamiltonianDecay} as

\begin{equation}\label{eq:PiI}
\begin{split}
\hat{\Pi}_{j,C} + \hat{\Pi}_{j,T} 
 = \sum_{q \in \{0,1\} } 
\Big( \dyad{r_j q}{r_j q} + \dyad{q r_j}{q r_j } \Big) + \sum^{N_{\mathrm{Ryd}}}_{k=1} \, \Big( \dyad{r_j r_k}{r_j r_k} + \dyad{r_k r_j}{r_k r_j} \Big) \, ,
\end{split}
\end{equation}
where we have used the resolution of identity $\mathbb{I} \!=\! \sum_{q \in \{0,1\}} \dyad{q}{q} + \sum^{N_{\mathrm{Ryd}}}_{k=1} \dyad{r_k}{r_k}$ for both the control and target atoms. Substituting Eq.~\eqref{eq:PiI} into the definition of $\hat{H}_{\mathrm{NH}}$ (Eq.~\eqref{eq:HamiltonianDecay} in the main text) and inserting the result into Eq.~\eqref{eq:UDecayO1}, we obtain

\begin{equation}\label{eq:UDecaymn1}
\begin{split}
    \hat{U}_{\mathrm{Decay,I}}(T) - \mathbb{I} 
    & \simeq 
    - \sum_{q \in \{0,1\} } \sum^{N_{\mathrm{Ryd}}}_{j=1} \, \frac{\Gamma_j}{2} \, \int^{T}_{0} dt \, \hat{U}^{\dagger}_{\mathrm{tot}}(t) 
\Big( \dyad{r_j q}{r_j q} + \dyad{q r_j}{q r_j } \Big) \hat{U}_{\mathrm{tot}}(t) \\
& - \sum^{N_{\mathrm{Ryd}}}_{j=1}\sum^{N_{\mathrm{Ryd}}}_{k=1} \, \frac{1}{2} \Bigg( \frac{\Gamma_j + \Gamma_k}{2} \Bigg) \, \int^{T}_{0} dt \, \hat{U}_{\mathrm{tot}}^{\dagger}(t) \, \Big( \dyad{r_j r_k}{r_j r_k} + \dyad{r_k r_j}{r_k r_j} \Big) \hat{U}_{\mathrm{tot}}(t)\, ,
\end{split}
\end{equation}
where in the second line, we have exploited the symmetry of the expression under $j \leftrightarrow k$ to replace $\Gamma_j$ with $(\Gamma_{j}+\Gamma_{k})/2$.

We next evaluate the trace over the computational states $\ket{01}$, $\ket{10}$, and $\ket{11}$, noting that $\ket{00}$ is fully decoupled from the dynamics, so that $(1/d)\,\mathrm{Tr}(\cdot) \!=\! (1/d) \sum_{p\in\{01,10,11\}} \bra{p} \cdot \ket{p}$. Combining this with Eqs.~\eqref{eq:UDecaymn1} and \eqref{eq:deltagDecayU}, and invoking the definitions of the cumulative losses in Eq.~\eqref{eq:Lambdas} of the main text, we obtain, to first order in $\Gamma_j T$,

\begin{equation}\label{eq:deltagDecay2}
\begin{split}
    \delta g_{\mathrm{Decay}} \simeq 
     -\sum_{q =0,1 } \! \sum^{N_{\mathrm{Ryd}}}_{j=1} \frac{1}{d}\sum_{p\in\{01,10,11\}} \Big[  \Lambda_{jq|p}(t) + (j \leftrightarrow q) \Big]  - \sum^{N_{\mathrm{Ryd}}}_{j,k=1} \frac{1}{d} \sum_{p\in\{01,10,11\}} \Big[  \Lambda_{jk|p}(t) + (j \leftrightarrow k) \Big] \, .
\end{split}
\end{equation}
Finally, using $\delta \varepsilon_{\mathrm{Decay}} = - 2 \delta g_{\mathrm{Decay}}$, we arrive at Eq.~\eqref{eq:dgsimple} of the main text.

\section{Exact solution of the dark state gate}
\label{supmatsec:ExactSolution}

\begin{figure}
    \centering
    \includegraphics[width=\textwidth]{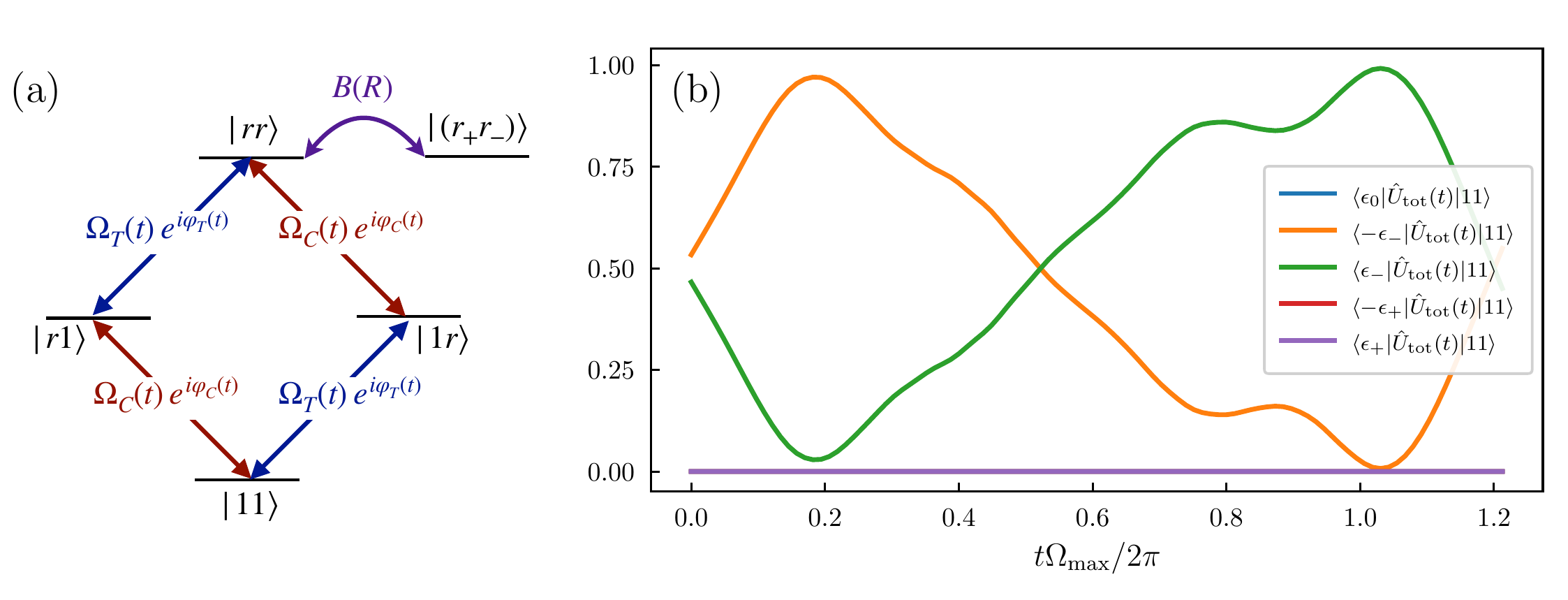}
    \caption{(a) Level diagram of states involved in the dynamics of the dark state gate connected to the $\ket{11}$ state. The zero-energy state $\ket{\epsilon_0}$ is decoupled from the dynamics at all times. The states $\ket{\pm \epsilon_{-}}$ have energies $\pm \epsilon_{-} \!\sim\! \Omega_{\mathrm{max}}$, while $\ket{\pm \epsilon_{+}}$ have energies $\pm \epsilon_{+} \!\sim\! B$ and are far off-resonant. (b) Time evolution of the exact eigenstate populations during the non-adiabatic gate dynamics from the initial state $\ket{11}$.}
    \label{fig:exactSolution}
\end{figure}

In this section, we derive the analytical eigenstates and eigenenergies that govern the dark state gate dynamics in the subspace coupled to $\ket{11}$, working in the limit of perfect Förster resonance where the van der Waals interaction $V$ is much smaller than both $B$ and the Rabi frequencies $\Omega_{C}$ and $\Omega_{T}$ of the control and target atoms. The resulting level scheme, coupled by the Rabi pulses, is shown in Fig.~\ref{fig:exactSolution}. This limit is instructive: it yields closed-form expressions for the eigenstates and eigenenergies, which in turn clarify the structure of the gate dynamics.

The regime $\Omega_{C},\Omega_{T} \!\ll\! B$ is of particular interest. A straightforward Taylor expansion in $\Omega_{C}/B$ and $\Omega_{T}/B$ shows that the leading-order correction to the eigenenergies and eigenstates from finite $B$ appears at second order in $\Omega_{C(T)}/B$. This second-order scaling underlies the reduced sensitivity of the dark state gate fidelity to variations in interaction strength, and hence to interatomic distance fluctuations.

Defining $\mathcal{B} \!=\! \big( B^4 + \Omega_C^2 \Omega_T^2 \big)^{1/4}$, the eigenenergies are $\{ \epsilon_0\!=\!0,\, \pm \epsilon_{-},\pm\epsilon_{+} \}$, with $\epsilon_{\pm} = (1/2)\sqrt{\left(2B^{2} + \Omega_{C}^{2} + \Omega_{T}^{2}\right)\pm 2\mathcal{B}^2}$. The corresponding normalized eigenstates in the basis $\{\ket{11},\ket{1r},\ket{r1},\ket{rr},\ket{(r_{+}r_{-})}\}$ are
\begin{equation}\label{eq:eigenstates}
\left\{
    \begin{array}{l}
    \ket{\epsilon_0} = \frac{1}{\mathcal{N}_0} \, 
    \begin{bmatrix}
0,\;
- e^{-i\varphi_{C}} \,\Omega_{C},\;
e^{-i\varphi_{T}} \,\Omega_{T},\;
0,\;
\frac{\Omega_{C}^{2} - \Omega_{T}^{2}}{2B}
    \end{bmatrix}^{T} \, ,
\\
    \ket{-\epsilon_{-}} = \frac{1}{\mathcal{N}_{-}} \, 
    \begin{bmatrix}
\frac{\epsilon_{-}\left(B^{2} + \mathcal{B}^{2}\right)}{B\,\Omega_{C}\Omega_{T}},\;
-\frac{e^{-i\varphi_{C}}\!\left(B^{2} + \mathcal{B}^{2} - \Omega_{C}^{2}\right)}
     {2B\,\Omega_{C}},\;
-\frac{e^{-i\varphi_{T}}\!\left(B^{2} + \mathcal{B}^{2} - \Omega_{T}^{2}\right)}
     {2B\,\Omega_{T}},\;
-\frac{\epsilon_{-}}{B},\;
1
    \end{bmatrix}^T \, ,
\\
    \ket{+\epsilon_{-}} = \frac{1}{\mathcal{N}_{-}} \, 
    \begin{bmatrix}
-\frac{\epsilon_{-}\!\left(B^{2} + \mathcal{B}^{2}\right)}{B\,\Omega_{C}\Omega_{T}},\;
-\frac{e^{-i\varphi_{C}}\!\left(B^{2} + \mathcal{B}^{2} - \Omega_{C}^{2}\right)}
     {2B\,\Omega_{C}},\;
-\frac{e^{-i\varphi_{T}}\!\left(B^{2} + \mathcal{B}^{2} - \Omega_{T}^{2}\right)}
     {2B\,\Omega_{T}},\;
\frac{\epsilon_{-}}{B},\;
1
    \end{bmatrix}^{T} \, ,
\\
    \ket{-\epsilon_{+}} = \frac{1}{\mathcal{N}_{+}} \, 
    \begin{bmatrix}
\frac{\epsilon_{+}\!\left(B^{2}-\mathcal{B}^{2}\right)}{B\,\Omega_{C}\Omega_{T}},\;
-\frac{e^{-i\varphi_{C}}\!\left(B^{2}-\mathcal{B}^{2}-\Omega_{C}^{2}\right)}
     {2B\,\Omega_{C}},\;
-\frac{e^{-i\varphi_{T}}\!\left(B^{2}-\mathcal{B}^{2}-\Omega_{T}^{2}\right)}
     {2B\,\Omega_{T}},\;
-\frac{\epsilon_{+}}{B},\;
1
    \end{bmatrix}^{T} \, ,
\\
    \ket{+\epsilon_{+}} = \frac{1}{\mathcal{N}_{+}} \, 
    \begin{bmatrix}
-\frac{\epsilon_{+}\!\left(B^{2}-\mathcal{B}^{2}\right)}{B\,\Omega_{C}\Omega_{T}},\;
-\frac{e^{-i\varphi_{C}}\!\left(B^{2}-\mathcal{B}^{2}-\Omega_{C}^{2}\right)}
     {2B\,\Omega_{C}},\;
-\frac{e^{-i\varphi_{T}}\!\left(B^{2}-\mathcal{B}^{2}-\Omega_{T}^{2}\right)}
     {2B\,\Omega_{T}},\;
\frac{\epsilon_{+}}{B},\;
1
    \end{bmatrix}^{T} \, ,
    \end{array}
\right.
\end{equation}
with norms
\begin{equation}\label{eq:norms}
\left\{
\begin{array}{l}
\mathcal{N}_{0}= \sqrt{ \left(\Omega_{C}^{2} + \Omega_{T}^{2}\right) + \frac{\left(\Omega_{C}^{2} - \Omega_{T}^{2}\right)^{2}}{4B^2} } \, ,
\\
\mathcal{N}_{\pm} \!=\! \sqrt{
\frac{(\Omega_{C}^{2} + \Omega_{T}^{2}) (B^2 \mp \mathcal{B}^2)}
     {\Omega_{C}^{2} \Omega_{T}^{2}}
+
\frac{(\Omega_{C}^{2} + \Omega_{T}^{2}) \pm 2 \mathcal{B}^2}
     {B^{2}}} \, .
    \end{array}
    \right.
\end{equation}
For the non-adiabatic gate dynamics discussed in the main text, $\Omega_{C}(t) \!=\! \Omega_T(t) \!=\! \Omega_{\mathrm{max}}$ and $\varphi_{C}(t) \!=\! \varphi_{T}(t) \!=\! \varphi(t)$. In this case, the eigenenergies reduce to
\begin{equation}\label{eq:bar-eps-pm}
    \bar{\epsilon}_{\pm} = \frac{1}{\sqrt{2}} \sqrt{B^2 + \Omega_{\mathrm{max}}^2 \pm \sqrt{B^4 + \Omega_{\mathrm{max}}^4 } } \, ,
\end{equation}
and the instantaneous eigenstates of Eq.~\eqref{eq:eigenstates} become

\begin{equation}\label{eq:eigenstatesEqualAmplitude}
\left\{
    \begin{array}{l}
    \ket{\epsilon_0} = \frac{e^{-i\varphi}}{\sqrt{2}} \, 
    \begin{bmatrix}
0,\;
-1 ,\;
1,\;
0,\;
0\end{bmatrix}^{T} \, ,
\\
    \ket{-\bar{\epsilon}_{-}} = \frac{1}{\bar{\mathcal{N}}_{-}} \, 
    \begin{bmatrix}
\frac{ \bar{\epsilon}_{-}\left( B^{2} + \sqrt{B^2 + \Omega^2_{\mathrm{max}}} \right)}{B \Omega^2_{\mathrm{max} } },\;
-\frac{e^{-i\varphi}\!\left( B^{2} + \sqrt{B^2 + \Omega^2_{\mathrm{max}}} - \Omega_{\mathrm{max}}^{2}\right)}
     {2B\,\Omega_{\mathrm{max}}},\;
-\frac{e^{-i\varphi}\!\left( B^{2} + \sqrt{B^2 + \Omega^2_{\mathrm{max}}} - \Omega_{\mathrm{max}}^{2} \right)}
     {2B\,\Omega_{\mathrm{max}} },\;
-\frac{ \bar{\epsilon}_{-}}{B},\;
1
    \end{bmatrix}^T \, ,
\\
\ket{+\bar{\epsilon}_{-}} = \frac{1}{\bar{\mathcal{N}}_{-}} \, 
    \begin{bmatrix}-
\frac{ \bar{\epsilon}_{-}\left( B^{2} + \sqrt{B^2 + \Omega^2_{\mathrm{max}}} \right)}{B \Omega^2_{\mathrm{max} } },\;
-\frac{e^{-i\varphi}\!\left( B^{2} + \sqrt{B^2 + \Omega^2_{\mathrm{max}}} - \Omega_{\mathrm{max}}^{2}\right)}
     {2B\,\Omega_{\mathrm{max}}},\;
-\frac{e^{-i\varphi}\!\left( B^{2} + \sqrt{B^2 + \Omega^2_{\mathrm{max}}} - \Omega_{\mathrm{max}}^{2} \right)}
     {2B\,\Omega_{\mathrm{max}} },\;
\frac{ \bar{\epsilon}_{-}}{B},\;
1
    \end{bmatrix}^T \, ,
\\
    \ket{-\bar{\epsilon}_{+}} = \frac{1}{\bar{\mathcal{N}}_{+}} \, 
    \begin{bmatrix}
\frac{\bar{\epsilon}_{+}\!\left( B^{2}-\sqrt{B^4+\Omega^4_{\mathrm{max}}} \right)}{B\,\Omega^2_{\mathrm{max}} },\;
-\frac{e^{-i\varphi}\!\left(B^{2}-\sqrt{B^4+\Omega^4_{\mathrm{max}}}-\Omega_{\mathrm{max}}^{2}\right)}
     {2B\Omega_{\mathrm{max}}},\;
-\frac{e^{-i\varphi}\!\left(B^{2}-\sqrt{B^4+\Omega^4_{\mathrm{max}}}-\Omega_{\mathrm{max}}^{2}\right)}
     {2B\Omega_{\mathrm{max}}},\;
-\frac{\bar{\epsilon}_{+}}{B},\;
1
    \end{bmatrix}^{T} \, ,
\\
    \ket{+\bar{\epsilon}_{+}} = \frac{1}{\bar{\mathcal{N}}_{+}} \, 
    \begin{bmatrix}
-\frac{\bar{\epsilon}_{+}\!\left(B^{2}-\sqrt{B^4+\Omega^4_{\mathrm{max}}}\right)}{B\,\Omega_{\mathrm{max}}^2},\;
-\frac{e^{-i\varphi}\!\left(B^{2}-\sqrt{B^4+\Omega^4_{\mathrm{max}}}-\Omega_{\mathrm{max}}^{2}\right)}
     {2B\,\Omega_{\mathrm{max}}},\;
-\frac{e^{-i\varphi}\!\left(B^{2}-\sqrt{B^4+\Omega^4_{\mathrm{max}}}-\Omega_{\mathrm{max}}^{2}\right)}
     {2B\,\Omega_{\mathrm{max}}},\;
\frac{\bar{\epsilon}_{+}}{B},\;
1
    \end{bmatrix}^{T} \, ,
    \end{array}
\right.
\end{equation}
where the norms $\mathcal{N}_{\pm}$ in Eq.~\ref{eq:norms} reduce to
\begin{equation}\label{eq:norm-pm-bar}
    \bar{\mathcal{N}}_{\pm} \!=\! \sqrt{ \frac{2 \big( B^2 \mp \sqrt{B^4+\Omega^4_{\mathrm{max}}} \big) }{\Omega^2_{\mathrm{max}}} + \frac{2\Omega^2_{\mathrm{max}} \pm 2 \sqrt{B^4+\Omega^4_{\mathrm{max}}} }{B^2} } \, .
\end{equation}

To illustrate the role of each eigenstate in the non-adiabatic gate dynamics, the exact eigenstate populations starting from the initial state $\ket{11}$ are shown in Fig.~\ref{fig:exactSolution}(b). It is worth emphasizing that when $\Omega_{C} \!=\! 0,\, \Omega_{T} \! \neq \! 0$ or $\Omega_{T} \!=\! 0,\, \Omega_{C} \! \neq \! 0$, the level scheme reduces to that of the (adiabatic) dark state gate: $\ket{\pm \epsilon_{-}}$ become the single-Rydberg-excitation states coupled by the nonzero Rabi pulse, $\ket{\epsilon_0}$ reduces to the dark state, and $\ket{\pm \epsilon_{+}}$ become far-off-resonant states.

\section{Fidelity error due to imperfections in Förster resonance tuning}
\label{sec:Foerstertuning}

The analysis in the main text assumes the resonance condition $\Delta \!=\! 0$, under which the Förster defect vanishes. In this section, we consider two experimental techniques for achieving this condition, namely, DC- and AC-Stark tuning. We then investigate the gate errors induced by experimental noise in each of these techniques. We find that AC-Stark tuning provides a substantially more robust route to achieving the resonance condition than DC-Stark tuning. These results demonstrate that the choice of Förster-tuning technique can have a significant impact on gate performance.

\subsection{DC-Stark tuning}

In this method, a constant electric field is applied along the quantization axis to bring the Rydberg pair states into resonance. For a specific electric field $E_{\mathrm{res}}$, the resonance condition $\Delta(E_{\mathrm{res}}) \!=\! 0$ can be achieved. In practice, this resonance is sensitive to stray electric fields and surface charges in glass-cell experiments, since UV Rydberg excitation leads to a finite F\"orster defect and a broadening of the resonance~\cite{mamat2024mitigating}. Recent electric-field cancellation techniques can suppress electric-field fluctuations to the level of $10\,\mathrm{mV/cm}$~\cite{panja2024electric}.

To quantify the impact of residual electric field fluctuations on the gate fidelity, we model the noise as shot-to-shot fluctuations that remain constant during a single gate operation. The resulting error has two contributions: shot-to-shot fluctuations in the ground--Rydberg transition frequency, and shot-to-shot fluctuations in the F\"orster defect.

We first consider the fluctuations in the F\"orster defect. For fluctuations $\delta E$ with $|\delta E| \lesssim 10\,\mathrm{mV/cm}$, the F\"orster defect can be accurately expanded to second order in $\delta E$ as
\begin{equation}
\Delta (E_{\mathrm{res}}+\delta E) = 
\left.\frac{d\Delta}{d E}\right|_{E=E_{\mathrm{res}}} \delta E 
+ \frac{1}{2} \left.\frac{d^2\Delta}{d E^2}\right|_{E=E_{\mathrm{res}}} \delta E^2 \, .
\end{equation}
The resonance field $E_{\mathrm{res}}$, together with the first and second derivatives of the F\"orster defect evaluated at $E_{\mathrm{res}}$, are listed in Table~\ref{tab:foersterresonanceparams} for different principal quantum numbers $n$, computed using the \texttt{PairInteraction} package.

\begin{figure}
    \centering
    \includegraphics[width=\textwidth]{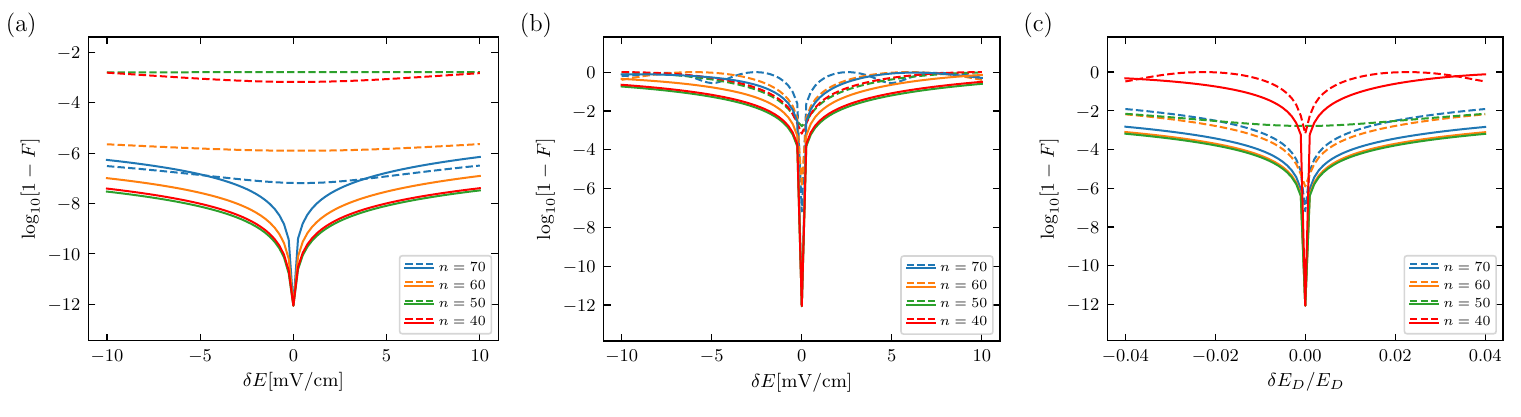} 
    \caption{(a) Fidelity error of the non-adiabatic (solid lines) and adiabatic (dashed lines) dark state gates due solely to F\"orster-defect fluctuations under DC-Stark tuning, shown for principal quantum numbers $n = 40$--$70$. (b) Fidelity error caused by combined fluctuations in the F\"orster defect and in the Rydberg transition frequency under DC-Stark tuning. (c) Same as (b), but under AC-Stark tuning.}
    \label{fig:FD_infidelity}
\end{figure}

The resulting gate infidelities due solely to F\"orster defect fluctuations, as a function of the shot-to-shot electric field noise, are shown in Fig.~\ref{fig:FD_infidelity}(a) for both the non-adiabatic (solid lines) and adiabatic (dashed lines) dark state gates, for principal quantum numbers in the range $n = 40$--$70$. The results show that F\"orster-defect fluctuations alone do not severely affect the gate infidelity. However, electric field fluctuations also cause fluctuations in the ground--Rydberg transition frequency. Figure~\ref{fig:FD_infidelity}(b) depicts the infidelities resulting from the combined effect of the Förster defect and the Rydberg transition frequency fluctuations. The results show that ground-Rydberg transition frequency fluctuations substantially increase the gate infidelities. However, the non-adiabatic gate exhibits greater robustness than the adiabatic gate: even for the largest principal quantum number considered, corresponding to the highest sensitivity, its infidelity remains below $10^{-6}$.

Figure~\ref{fig:FD_infidelity}(a) shows the gate infidelities arising solely from Förster-defect fluctuations, as a function of shot-to-shot electric field noise, for both the non-adiabatic (solid lines) and adiabatic (dashed lines) dark state gates and principal quantum numbers $n = 40$--$70$. These results show that Förster defect fluctuations alone do not substantially affect the gate infidelity. However, electric field fluctuations also shift the ground--Rydberg transition frequency. Figure~\ref{fig:FD_infidelity}(b) shows the infidelities resulting from the combined effect of Förster defect and ground--Rydberg transition frequency fluctuations. Here, the transition frequency fluctuations substantially increase the gate infidelities, revealing the sensitivity of the gate fidelity to electric-field fluctuations under DC-Stark tuning. In the following, we show that AC-Stark tuning yields a much smaller error budget, offering a more reliable route to implementing the dark state gate.

\begin{table}[h]
    \centering
    \begin{tabular}{| c || c | c | c |}
        \hline
        $n$ & $E_{\mathrm{res}}$ (V/cm) & $d\Delta/dE|_{E=E_{\mathrm{res}}}\, ( \mathrm{MHz} \cdot \mathrm{cm} /\mathrm{mV})$  & $d^{2}\Delta/dE^2|_{E=E_{\mathrm{res}}} \, ( \mathrm{MHz} \cdot \mathrm{cm}^2 /\mathrm{mV}^2)$ \\
        \hline
        40 & 1.4547 & 0.55128 & $3.54 \times 10^{-4}$ \\
        \hline
        50 & 0.228 & 0.48938 & $2.08 \times 10^{-3}$ \\
        \hline
        60 & 0.1115 & 0.92596 & $8.02 \times 10^{-3}$ \\
        \hline
        70 & 0.0862 & 2.17143 & $2.29 \times 10^{-2}$ \\
        \hline
    \end{tabular}
    \caption{
    Dependence on the principal quantum number of the resonance electric field $E_{\mathrm{res}}$, together with the first and second derivatives of the F\"orster defect with respect to the electric field, evaluated at $E_{\mathrm{res}}$ and denoted by $d\Delta/dE|_{E=E_{\mathrm{res}}}$ and $d^{2}\Delta/dE^2|_{E=E_{\mathrm{res}}}$, respectively.
    }
    \label{tab:foersterresonanceparams}
\end{table}

\subsection{AC-Stark tuning}

In AC-Stark tuning, the differential AC-Stark shifts of the states involved in the F\"orster resonance are used to bring the initial and final pair states into resonance. We consider a $\pi$-polarized drive at frequency $\omega_D$ with electric field amplitude $E_D$ such that the drive field is $\vc{E}_D(t) \!=\! \vc{e}_{z} \, (E_D/2) \big( e^{i\omega_D t} + e^{-i\omega_D t} \big)$. The resulting AC-Stark shift of a Rydberg state $\ket{i}$ is given by $\delta E^{\mathrm{st}}_{i}(\omega_D,E_D) \!=\! -\tfrac{1}{4} \, \alpha_{i}(\omega_D) \,E_D^2$, where
\begin{equation}\label{eq:dynpol}
    \alpha_{i}(\omega) = \sum_{k \neq i} \, |\bra{k} \hat{d}_{z} \ket{i} |^2 \bigg[ \frac{1}{\omega_{k} - \omega_{i} - \omega } + \frac{1}{\omega_{k} - \omega_{i} + \omega } \bigg]
\end{equation}
is the dynamic polarizability~\cite{cohen2024atom}, the states $\ket{k}$ are the off-resonantly coupled single-atom states with energies $\omega_k$, and $\hat{d}_z$ is the dipole operator along the quantization axis. The resulting shift of the F\"orster defect is
\begin{equation}
    \delta \Delta (\omega_D,E_D) = \delta E^{\mathrm{st}}_{r_{+}}(\omega_D,E_D) + \delta E^{\mathrm{st}}_{r_{-}}(\omega_D,E_D) - 2\, \delta E^{\mathrm{st}}_{r}(\omega_D,E_D) \, .
\end{equation}
The AC-Stark drive frequency and amplitude can be tuned to values $\omega^*_D$ and $E^*_D$ satisfying $\Delta_0 + \delta \Delta(\omega^*_D,E^*_D) \!=\! 0$, where $\Delta_0$ is the natural F\"orster defect in the absence of Stark tuning. The state $\ket{r}$ is simultaneously shifted by $\delta E^{\mathrm{st}}_{r}(\omega^{*}_D,E^{*}_D)$, an effect that can be compensated by adjusting the Rydberg excitation laser frequency accordingly.

To evaluate the effect of AC electric field fluctuations on the gate fidelity, we neglect fluctuations in the drive frequency and consider only shot-to-shot fluctuations in the drive electric field amplitude $E_D$. Under this assumption, at each gate operation, the Rydberg state $\ket{r}$ is shifted to $\delta E^{\mathrm{st}}_{r}(\omega^*_D,E_D)$ and the Förster defect to $\delta \Delta(\omega^*_{D},E_D)$, which we expand to first order in $E_D - E_D^*$ as
\begin{equation}
\begin{split}
    & \delta \Delta(\omega^*_{D},E_D) \simeq \delta \Delta(\omega^{*}_{D},E^{*}_D) + 
    \frac{ d  \delta \Delta}{d E_D} \Bigg|_{\omega^*_D,E^*_D} \big(E_D - E^*_D \big) \, ,\\
    & \delta E^{\mathrm{st}}_{r}(\omega^*_D,E_D) \simeq \delta E^{\mathrm{st}}_{r}(\omega^*_D,E^*_D) + \frac{d \delta E^{\mathrm{st}}_{r}}{d E_D} \bigg|_{\omega^*_D,E^*_D} \big(E_D - E^*_D \big) \, .
\end{split}
\end{equation}
These fluctuations in $\delta \Delta$ and $\delta E^{\mathrm{st}}_r$ translate directly into errors in the gate fidelity. The resulting gate infidelities due to fluctuations of the AC-Stark drive amplitude are shown in Fig.~\ref{fig:FD_infidelity}(c), for both the adiabatic and non-adiabatic dark state gates and principal quantum numbers $n \!=\! 40$--$70$. Comparing panels (b) and (c) of Fig.~\ref{fig:FD_infidelity} shows that AC-Stark tuning yields significantly smaller gate errors than DC-Stark tuning, offering a more reliable and robust route to implementing the non-adiabatic dark state gate. The parameters used to calculate the gate infidelities are obtained using the $\mathtt{ARC}$ package~\cite{vsibalic2017arc} and are listed in Table~\ref{tab:ac-stark-params}. 

\begin{table}[h]
    \centering
    \begin{tabular}{| c || c | c || c | c|}
        \hline
        $n$ & $\omega^{*}_{D}/2\pi$ (GHz) & $E^{*}_{D}$ (V/cm) & $1/2\pi \cdot d \delta \Delta/dE_D|_{\omega^*_D,E^*_D}$ (MHz$\cdot$cm/V) & $ 1/2\pi \cdot d\delta E^{\mathrm{st}}_{r}/dE_D|_{\omega^*_D,E^*_D}$ (MHz$\cdot$cm/V)\\
        \hline
        40 & 11.20 & 0.30 & 843.46 & -423.59 \\
        \hline
        50 & 28.50 & 0.33 & -1353.93 & 10.16 \\
        \hline
        60 & 16.0 & 0.18 & -2586.82 & 21.08 \\
        \hline
        70 & 9.80 & 0.12 & -3170.30 & 43.79 \\
        \hline
    \end{tabular}
    \caption{AC-Stark drive frequency and amplitude required to achieve the resonance condition $\Delta_0 + \delta \Delta(\omega^*_D,E^*_D) \!=\! 0$ for principal quantum numbers $n = 40$--$70$, together with the first derivatives of $\delta \Delta$ and $\delta E^{\mathrm{st}}_r$ with respect to the drive electric-field amplitude, evaluated at the operating point.
    }
    \label{tab:ac-stark-params}
\end{table}

\end{document}